\documentclass[prc,aps,amsmath,amssymb,superscriptaddress,twocolumn,showpacs,floatfix,a4paper]{revtex4-2}

\usepackage{graphicx,colordvi}
\usepackage{dcolumn}
\usepackage{bm}
\usepackage{threeparttable}
\usepackage{xspace}
\usepackage{gensymb}
\usepackage{cases}
\usepackage{tabularx,booktabs}
\usepackage{epstopdf}
\usepackage{xcolor}
\usepackage{physics} 
\usepackage{mathtools, amssymb, amsthm, amsmath}
\usepackage{physics} 
\usepackage{appendix}
\usepackage{csquotes}

\usepackage{xcolor}
\usepackage{tcolorbox}

\usepackage[dvipsnames]{xcolor}

\begin{document}

\title{Microscopic description of the fission process including intrinsic excitations \\
Part II: $^{240}$Pu excited and asymmetric fission paths within the Schr\"odinger Collective Intrinsic Model}

\author{P. Carpentier}
\affiliation{%
 CEA, DAM, DIF, F-91297 Arpajon cedex, France
}%
\affiliation{%
 Universit\'e Paris-Saclay, CEA, LMCE, 91680, Bruyères-le-Châtel, France
}%

\author{N. Pillet}%
\affiliation{%
 CEA, DAM, DIF, F-91297 Arpajon cedex, France
}%
\affiliation{%
Universit\'e Paris-Saclay, CEA, LMCE, 91680, Bruyères-le-Châtel, France
}%

\author{R. Bernard}%
\affiliation{%
CEA, DES, IRESNE, DER, SPRC, LEPh, 13115 Saint-Paul-lès-Durance, France
}%

\author{L.M. Robledo}%
\affiliation{%
Center for Computational Simulation, Universidad Polit\'ecnica de 
Madrid, Campus Montegancedo, 28660 Boadilla del Monte, Madrid, Spain
}%
\affiliation{Departamento  de F\'{\i}sica Te\'orica and CIAFF, 
Universidad Aut\'onoma de Madrid, 28049-Madrid, Spain}%

\author{D. Lacroix}%
\affiliation{%
Universit\'e Paris-Saclay, CNRS/IN2P3, IJCLab, Orsay, 91405, France
}%

\author{N. Dubray}%
\affiliation{%
 CEA, DAM, DIF, F-91297 Arpajon cedex, France
}%
\affiliation{%
Universit\'e Paris-Saclay, CEA, LMCE, 91680, Bruyères-le-Châtel, France
}%

\author{D. Regnier}%
\affiliation{%
 CEA, DAM, DIF, F-91297 Arpajon cedex, France
}%
\affiliation{%
 Universit\'e Paris-Saclay, CEA, LMCE, 91680, Bruyères-le-Châtel, France
}%

\author{W. Younes}%
\affiliation{%
 Nuclear Science Division, Lawrence Berkeley National Laboratory, Berkeley, California 94720, USA
}%

\date{\today}

\begin{abstract}
This second article of the trilogy \cite{trilogy1,trilogy2,trilogy3} presents the implementation of a third protocol, referred to as \enquote{Continuous Deflation}, designed to construct continuous and regular excited paths within the Schrödinger Collective-Intrinsic Model (SCIM), with applications to nuclear fission.
We show that the use of standard 2QP excitations, even when combined with particle-number projection, prevents a consistent application of the SCIM framework. Motivated by the central role of pair-breaking in low-energy fission, we explore how to construct intrinsic excited states that incorporate this mechanism while satisfying the continuity and regularity state requirements of the SCIM. To this end, we first analyze the \enquote{Deflation} procedure alone, which constructs excited states through orthogonality constraints. We then extend this construction along a deformation path by introducing an additional continuity constraint, thereby defining the \enquote{Continuous Deflation} method, which generates continuous paths based on excited states. In particular, we construct ten such continuous paths built on top of the adiabatic and asymmetric fission path of $^{240}$Pu. The resulting excited states are systematically analyzed in terms of their microscopic structure. We then investigate several fragment properties near scission, including neutron and proton chemical potentials, neutron necking as well as fragment particle-number distributions, and compare them with their adiabatic counterparts.

\end{abstract}

\maketitle

\section{Introduction}

In the first article of this trilogy \cite{trilogy1}, we discussed the importance of continuity and regularity of the states used for the Schrödinger Collective Intrinsic Model (SCIM). We introduced two new protocols, the \enquote{Link} and \enquote{Drop} methods \cite{carpentier2024,TPaul}, which rely on overlap constraints to generate one-dimensional continuous adiabatic paths with improved regularity properties. These methods enabled us to describe, in terms of Hartree-Fock-Bogoliubov (HFB) states, the static evolution of a fissioning compound nucleus from its ground state to and beyond scission along a one-dimensional adiabatic potential energy surface (PES).

In this second article, we investigate which classes of excited states satisfy the stringent continuity and regularity requirements imposed by the SCIM framework. In Section \ref{2QPdiscuss}, we examine the standard two-quasiparticle (2QP) excitations formulated within HFB theory, originally introduced in the SCIM framework to account for the low-energy pair-breaking mechanism \cite{bernard1}. We discuss their limitations and shortcomings in detail. In Section \ref{deflation}, we introduce a third protocol, referred to as \enquote{Continuous Deflation}, which complements the \enquote{Link} and \enquote{Drop} methods. Like its two predecessors, this protocol is based on overlap constraints and simultaneously enforces continuity and orthogonality, thereby enabling the construction of continuous and regular excited states expressed as HFB vacua. 
In Section~\ref{correlation}, we analyze the microscopic structure of the excited states generated by the \enquote{Continuous Deflation} method. Their continuity and regularity properties are also investigated and compared with those of the adiabatic states along the asymmetric fission path of $^{240}$Pu. Finally, Section~\ref{scissionarea} is devoted to the study of the properties
of such excited states in the scission area of $^{240}$Pu, such as the neutron necking and proton and neutron fragment distribution.
Section \ref{conclusion} presents conclusions and outlines future perspectives to this work.

\section{2QP and projected 2QP excited states issues }\label{2QPdiscuss}

In the original SCIM framework \cite{bernard1}, excited states were modeled as non-self-consistent 2QP excitations built on top of the HFB adiabatic states at each value of the collective coordinate. As the elementary excitations of HFB theory, quasiparticle (QP) states provide a natural starting point for describing excited configurations and inherently account for pair-breaking effects, which play a crucial role in low-energy fission. However, as will be shown below, their intrinsic limitations prevent the direct application of the SCIM formalism. 

To ensure consistency with the time-reversal-invariant adiabatic states of even-even nuclei, the original SCIM framework considers time-even 2QP excited states of the form:
\begin{eqnarray}
\ket{\Phi_{ij}} = \alpha_{ij}(\xi^+_i \bar \xi_j^+ + \xi^+_j \bar \xi_i^+)\ket{\Phi_{HFB}}
\end{eqnarray}
where the set of $\xi^+_l$ represents QP creation operators and the normalization constant $\alpha_{ij}$ is defined such that:
\begin{equation}
\begin{array}{lcl}
\bra{\Phi_{ij}}\ket{\Phi_{ij}} &=& 
\displaystyle 2\alpha_{ij}^2(1 + \delta_{ij}) = 1.
\end{array}
\end{equation}
To preserve axial symmetry, the QP $i$ and $j$ are chosen within the same $\Omega$ block, corresponding to a given projection of the angular momentum onto the $z$ axis. They are also required to have the same isospin projection $\tau$.

Since QP states are not uniquely determined by the quantum numbers $\Omega$ and $\tau$, their identification is intrinsically ambiguous. However, the SCIM formalism requires each excitation to be uniquely identified along the entire deformation path. To achieve this, the excited state $\ket{\Phi_{ij}(q')}$ at deformation $q' = q + \delta q$ is associated with the neighboring state $\ket{\Phi_{ij}(q)}$ at deformation $q$ by maximizing their overlap according to:
\begin{eqnarray}
 \displaystyle \vert \bra{\Phi_{ij}(q)}\ket{\Phi_{ij}(q')} \vert = \text{max}_{i'j'}\vert \bra{\Phi_{ij}(q)}\ket{\Phi_{i'j'}(q')} \vert.
\end{eqnarray}

The procedure used to evaluate these overlaps is described in detail in Appendix D of the first article of this trilogy \cite{trilogy1}.

\begin{figure}
\centering
\includegraphics[width=1.0\linewidth]{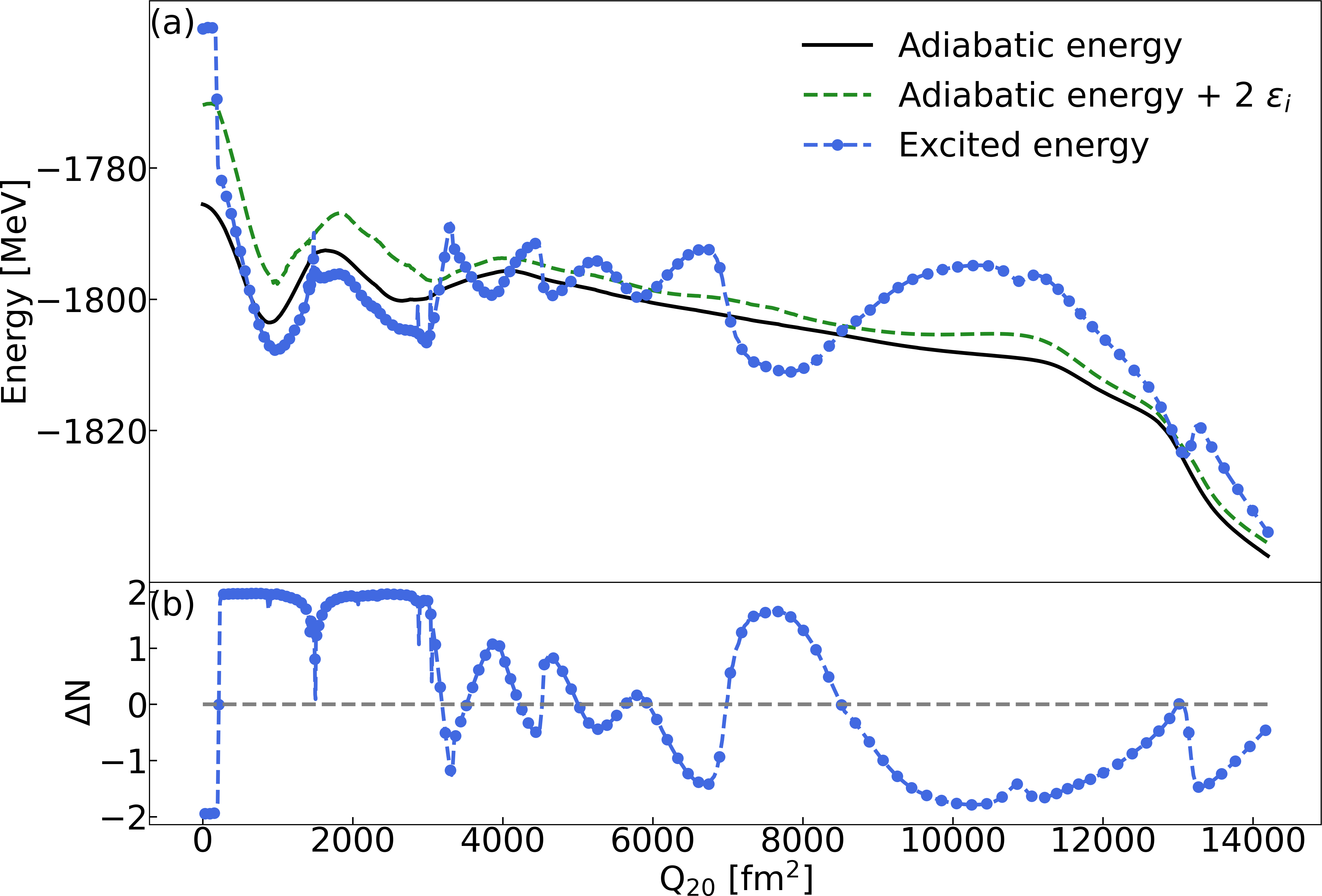}
\caption{Panel (a): Adiabatic potential energy surface obtained with the $\tilde{\mathcal{P}}_{20}$ procedure (black), compared with the energy of a $\Omega=1/2$ neutron 2QP excitation (blue) and with its HFB approximation augmented by twice the QP energy $\epsilon_i$ (green). The protocol to build the 2QP states is described in the text. Panel (b): Particle-number difference $\Delta N$ between the adiabatic and 2QP excited states. The horizontal line indicates $\Delta N = 0$. Calculations are performed for the asymmetric fission path of $^{240}$Pu.}
\label{fig_qp}
\end{figure}

A major limitation of non-self-consistent 2QP excited states is their poor conservation of the average particle number, which may vary significantly along the deformation path. As illustrated in FIG. \ref{fig_qp}, these particle-number fluctuations give rise to pronounced discontinuities in the excitation energy.

The comparison between the exact 2QP energy and its HFB approximation reveals that the two agree only when the particle-number difference $\Delta N$ between the excited and adiabatic states remains close to zero. As soon as $\Delta N$ increases, the exact excitation energy departs markedly from the HFB estimate, highlighting the strong influence of particle-number breaking on the Hamiltonian kernel.

This issue was already recognized in Ref.~\cite{bernard1}, where the proposed remedy consisted in selecting 2QP configurations that minimize particle-number fluctuations. However, FIG. \ref{fig_qp} shows that such a selection cannot generally be maintained over the entire deformation path. Moreover, even when $\Delta N$ remains within the interval $]-1,1[$, the residual particle-number fluctuations induce spurious variations of the Hamiltonian kernel. These artificial fluctuations significantly degrade the regularity of the kernels, a key requirement for reliable SCIM calculations. 

Since particle-number breaking is the main source of the deficiencies discussed above, a most natural way to address this issue is to restore particle number by projection. Two standard approaches are available: Projection After Variation (PAV) and Variation After Projection (VAP) \cite{PAngRobl}. In the present work, we adopt the PAV scheme, whose lower computational cost makes it well suited for assessing the impact of particle-number restoration within the SCIM framework.

For an HFB state $\ket{\Phi_{HFB}}$, the particle-number projected state is defined as:
\begin{equation}\label{ctwo_27}
\ket{\Phi_{HFB}^{Proj}} = \frac{ \hat P_{N_0}}{\sqrt{\mathcal{C}}} \ket{\Phi_{HFB}}
\end{equation}
where $\hat P_{N_0}$ denotes the projector onto the particle number $N_0$ \cite{PAngRobl}, and the normalization coefficient $\mathcal{C}$ is given by:
\begin{equation}
\mathcal{C} = \bra{\Phi_{HFB}} \hat P_{N_0} \ket{\Phi_{HFB}}
\end{equation}
ensures proper normalization.

The effect of particle-number projection on HFB states is well established \cite{PAngRobl}. By restoring particle number, the projection recovers part of the pairing correlation energy, leading to an increase in binding energy whose magnitude is closely related to the particle-number dispersion of the underlying HFB state.\\

\begin{figure}
\centering
\includegraphics[width=1.0\linewidth]{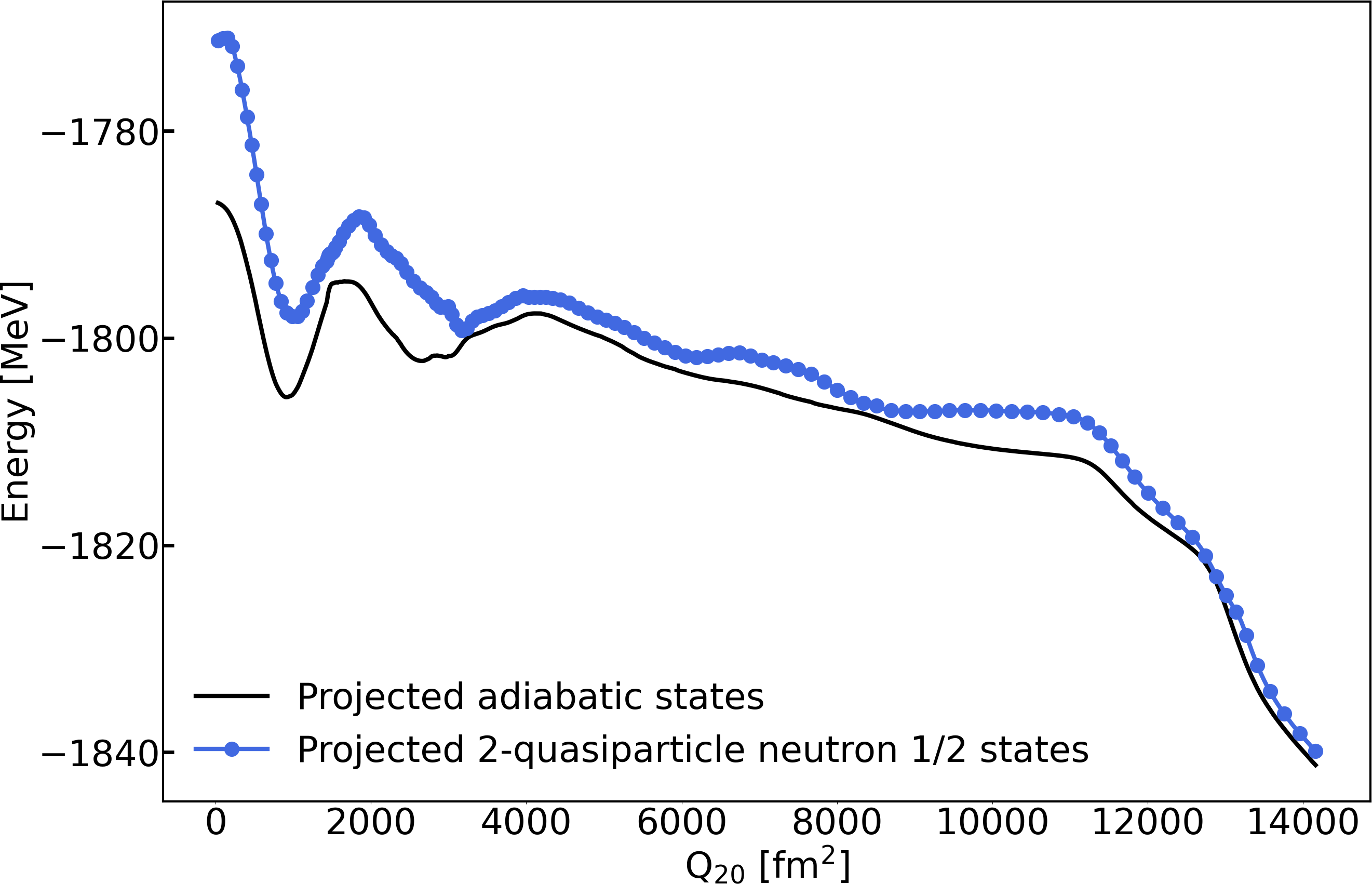}
\caption{Projected adiabatic potential energy surface (black) and projected neutron $\Omega=1/2$ 2QP excitation (blue) along the asymmetric fission path of $^{240}$Pu as a function of the quadrupole moment $Q_{20}$.}
\label{ctwo_28}
\end{figure}

By analogy with the definition of the projected adiabatic HFB state \eqref{ctwo_27}, the particle-number projected 2QP excited states are defined as:
\begin{eqnarray}
\ket{\Phi_{ij}^{Proj}} = \frac{\hat P_{N_0}}{\sqrt{\mathcal{C}_{ij}}}(\xi^+_i \bar \xi_j^+ + \xi^+_j \bar \xi_i^+)\ket{\Phi_{HFB}},
\end{eqnarray}
where the normalization constant is given by:
\begin{equation}
\mathcal{C}_{ij} = \bra{\Phi_{HFB}}(\bar \xi_i  \xi_j + \bar \xi_j  \xi_i)\hat P_{N_0}(\xi^+_i \bar \xi_j^+ + \xi^+_j \bar \xi_i^+)\ket{\Phi_{HFB}}.
\end{equation}
As expected, particle-number projection substantially improves the energy of the 2QP excited states. As shown in FIG. \ref{ctwo_28}, the projected excitation energy becomes smooth and remains above the projected adiabatic potential energy surface, in close agreement with the HFB estimate discussed previously. Particle-number restoration therefore successfully removes the spurious energy variations induced by particle-number breaking.

\begin{figure}
\centering
\includegraphics[width=1.0\linewidth]{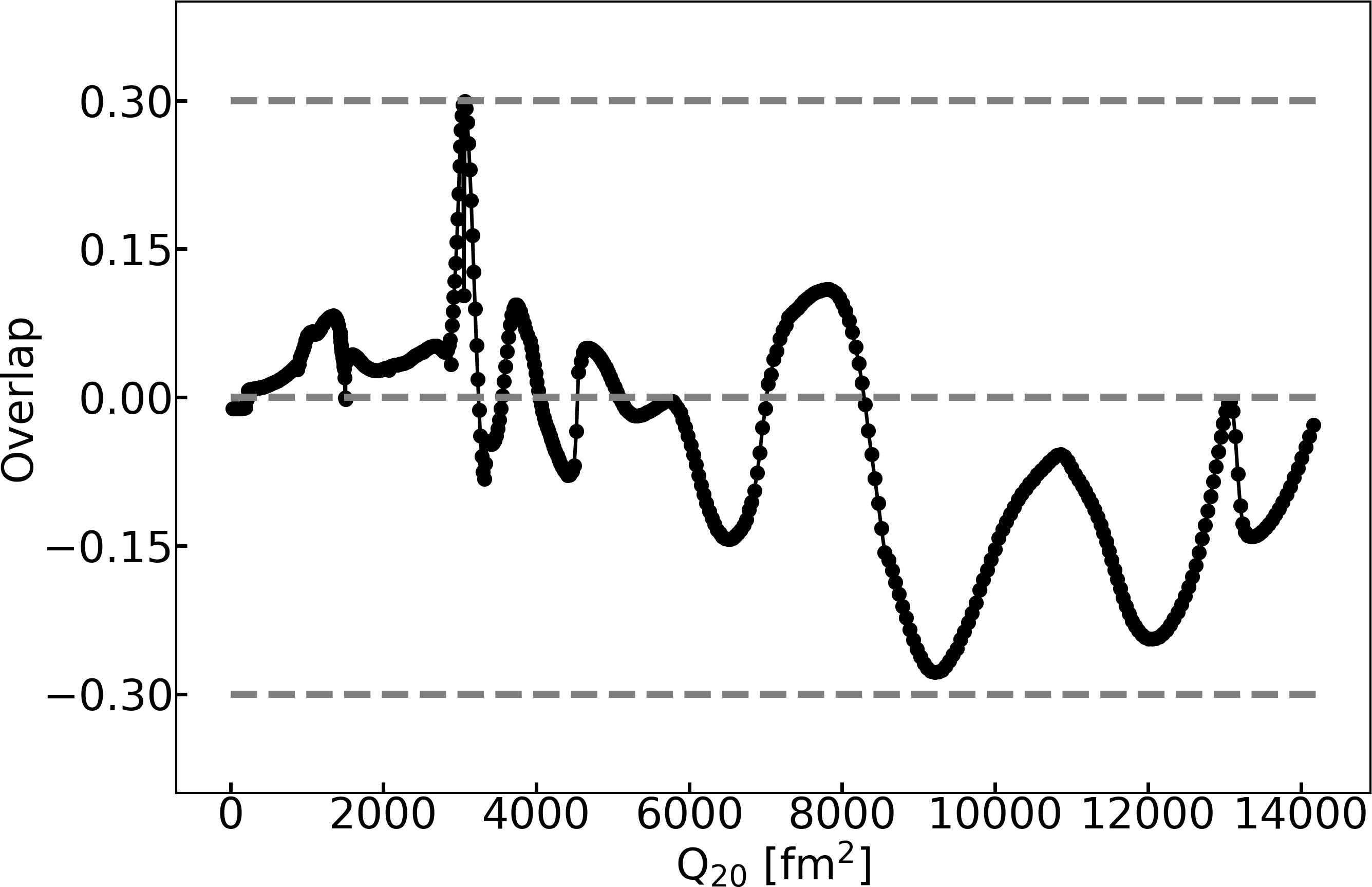}
\caption{Overlap between the projected adiabatic HFB state and the corresponding projected 2QP excited state along the asymmetric fission path of $^{240}$Pu.}
\label{ctwo_30}
\end{figure}

However, this improvement comes at the expense of another fundamental requirement of the SCIM framework. Indeed, the projected 2QP excited states are no longer orthogonal to the projected adiabatic HFB states. FIG. \ref{ctwo_30} shows that the overlap between these two classes of states is far from negligible and exhibits significant variations along the deformation path. Such a loss of orthogonality prevents a clear separation between collective and intrinsic degrees of freedom. As a result, the projected excited states cannot be regarded as genuine intrinsic excitations built on top of the adiabatic reference states, making their interpretation within the SCIM framework fundamentally ambiguous.

\begin{figure}
\centering
\includegraphics[width=1.0\linewidth]{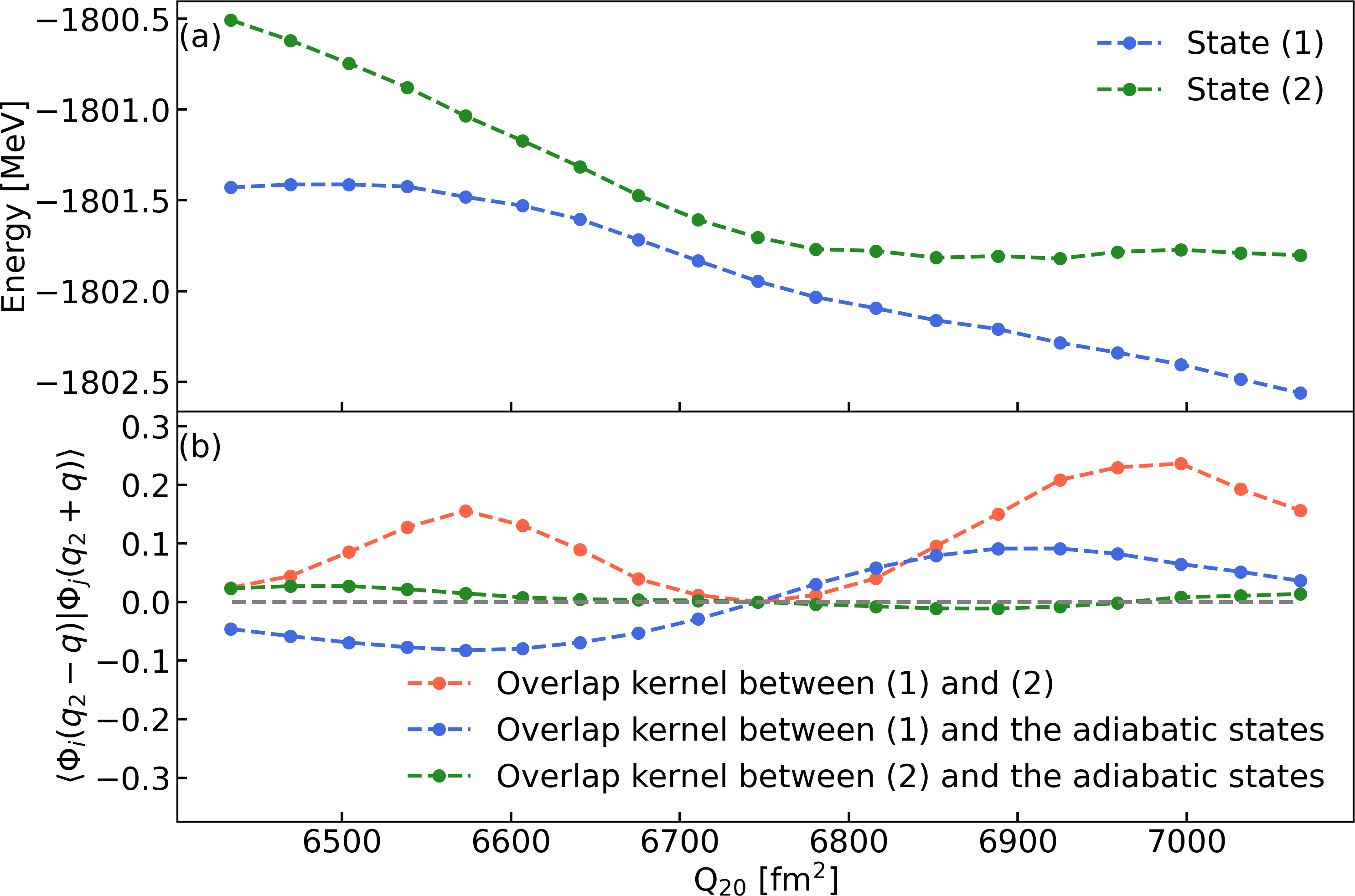}
\caption{Level repulsion between two $\Omega = 1/2$ 2QP excited states along the asymmetric fission path of $^{240}$Pu at $Q_{20} = q_2 \simeq 6750$ fm$^2$. Panel (a): Projected potential energy surfaces of the two involved configurations. Panel (b): Overlap kernels between the two excited states (red) and between each excited state and the corresponding adiabatic reference state (blue and green).}
\label{ctwo_50}
\end{figure}

The limitations of the PAV scheme could, in principle, be mitigated by adopting a Variation After Projection (VAP) approach, albeit at a significantly higher computational cost. However, both strategies ultimately face a more fundamental difficulty: level repulsions along the fission path, which prevent a consistent tracking of 2QP excitations within the SCIM framework.
As illustrated in FIG. \ref{ctwo_50}, where such a repulsion occurs around $Q_{20} \simeq 6750$ fm$^2$ for two $\Omega = 1/2$ 2QP states along the asymmetric fission path of $^{240}$Pu. In this region, the two excited configurations labeled by (1) and (2) undergo a strong mixing and effectively exchange their character, as evidenced by the behavior of the corresponding overlap kernels. This phenomenon reflects the absence of a smooth one-to-one correspondence between QP configurations along the deformation path. As the collective coordinate evolves, the underlying QP content is sometimes strongly reshuffled due to avoided crossings. Such rearrangements induce irregular variations in the excitation kernels. In an ideal formulation, adiabatic and excited configurations would be defined on distinct collective coordinates; however, the SCIM framework enforces a single shared collective coordinate, making such a separation impossible. As a consequence, our conclusion is that 2QP-based excited states are not suitable building blocks for a regular SCIM expansion. This motivates the alternative construction introduced in the next section.
\section{Deflation-based construction of excited states}\label{deflation}

\subsection{\enquote{Deflation} method}

In the context of the \enquote{Deflation} technique used in quantum computing \cite{merciDenis} and quantum chemistry \cite{QuantumDef}, excited variational states are constructed by enforcing orthogonality constraints with respect to previously determined states. In practice, this is achieved by augmenting the energy functional with projector-like penalty terms, leading to a constrained Hamiltonian $\hat{H}_c$ of the form:
\begin{eqnarray}\label{ctwo_56}
 \hat H_c = \hat H + \mu_N \hat{N} + \mu_Z \hat{Z} + \sum_\alpha \lambda_\alpha \hat Q_\alpha \nonumber\\ + \sum_\beta \gamma_\beta \ket{\Phi_\beta}\bra{\Phi_\beta},
\end{eqnarray}
\noindent where $\mu_N$ and $\mu_Z$ denote the neutron and proton chemical potentials, and the Lagrange multipliers $\lambda_\alpha$ enforce the usual constraints on multipole moments. The additional coefficients $\gamma_\beta$ are associated with orthogonality constraints imposed on a set of reference states $\ket{\Phi_\beta}$, as already introduced in the first article of this trilogy to ensure continuity along the deformation path.

In addition, the \enquote{Deflation} method relies on the iterative construction of excited states by minimizing the total energy (here, the HFB one) under orthogonality constraints with respect to previously determined configurations, while allowing for additional physical constraints.

In practice, orthogonality conditions can be imposed in a flexible manner, depending on the selected degrees of freedom. For a given reference state $\ket{\Phi_\beta}$, one may, for instance, enforce orthogonality in selected proton or neutron isospins, or restrict it to specific angular-momentum projection blocks $\Omega$, while leaving the remaining degrees of freedom unconstrained. Additional constraints on selected overlap components may also be included when required by the physical situation.

The general algorithm proceeds iteratively as follows. Starting from a reference adiabatic state $\ket{A}$, the first deflated state $\ket{D_1}$ is obtained by minimizing the HFB energy under the imposed constraints and orthogonality to $\ket{A}$. Higher excited states $\ket{D_i}$ are then constructed sequentially by enforcing orthogonality to all previously determined states ${\ket{D_j}}_{j<i}$, while maintaining the same set of physical constraints and minimizing the energy at each step.
In all calculations presented in this work, the particle numbers and the expectation value of the quadrupole operator $\hat{Q}_{10}$ are consistently constrained.

To provide further insight into this technique newly introduced in nuclear physics \cite{carpentier2024}, we illustrate the \enquote{Deflation} procedure through two representative cases. The first one focuses on the $^{16}$O nucleus, where the simplicity of the system allows us to analyze the microscopic structure generated by the \enquote{Deflation} method. We then consider the more complex case of the $^{240}$Pu nucleus, demonstrating the ability of the method to generate a set of mutually orthogonal variational excited states in a realistic nuclear system. Additional details can be found in Ref. \cite{TPaul}.

\subsubsection{Microscopic \enquote{Deflation} study in the $^{16}$O nucleus}

It is instructive to illustrate the microscopic action of the Deflation method on the simple case of the doubly magic nucleus $^{16}$O. At the HFB level, the ground state of the $^{16}$O nucleus is spherical and does not exhibit pairing correlations. Consequently, the canonical basis is characterized by occupation numbers equal to either 0 or 1. For each isospin, the occupied canonical space corresponds to the closed-shell configuration $1s^2_{1/2},1p^4_{3/2},1p^2_{1/2}$, yielding six occupied $\Omega = \pm 1/2$ states and two occupied $\Omega = \pm 3/2$ states.

When a \enquote{Deflation} is performed while imposing orthogonality only within a given isospin, a remarkable reorganization of the canonical basis takes place. The $\Omega=\pm1/2$ blocks remain strictly identical to those of the HFB ground state. By contrast, each of the $\Omega=3/2$ and $\Omega=-3/2$ blocks associated with the constrained isospin now contains two canonical states with occupation numbers equal to 0.5. The same behavior is obtained independently for neutrons and protons, reflecting the symmetry of the system.

From a physical point of view, given the symmetries imposed in the calculation and the closed-shell nature of the nucleus, one expects the lowest orthogonal variational configuration to resemble a particle-hole excitation, in which a nucleon is promoted from an occupied shell to a higher-lying unoccupied orbital. Such a picture, however, immediately raises the question of which member of a time-reversed pair should be excited. The present \enquote{Deflation} procedure naturally preserves time-reversal symmetry by treating the $\Omega=3/2$ and $\Omega=-3/2$ partners on an equal footing. The resulting variational state therefore distributes the occupation equally between the two members of the pair, leading to the four half-occupied canonical states observed above. At this level, the \enquote{Deflation} method therefore reproduces the expected microscopic mechanism of a particle-hole excitation while preserving the fundamental symmetries of the HFB framework.

Although this example corresponds to a particularly simple nucleus, it provides a clear and intuitive microscopic illustration of how the \enquote{Deflation} procedure modifies the underlying HFB structure to generate orthogonal configurations. We now turn to the considerably more challenging case of the $^{240}$Pu nucleus, where the method is used to construct a sequence of mutually orthogonal variational excited states.

\subsubsection{\enquote{Deflation} spectrum in the $^{240}$Pu nucleus}

\begin{figure}
\centering
\includegraphics[width=0.8\linewidth]{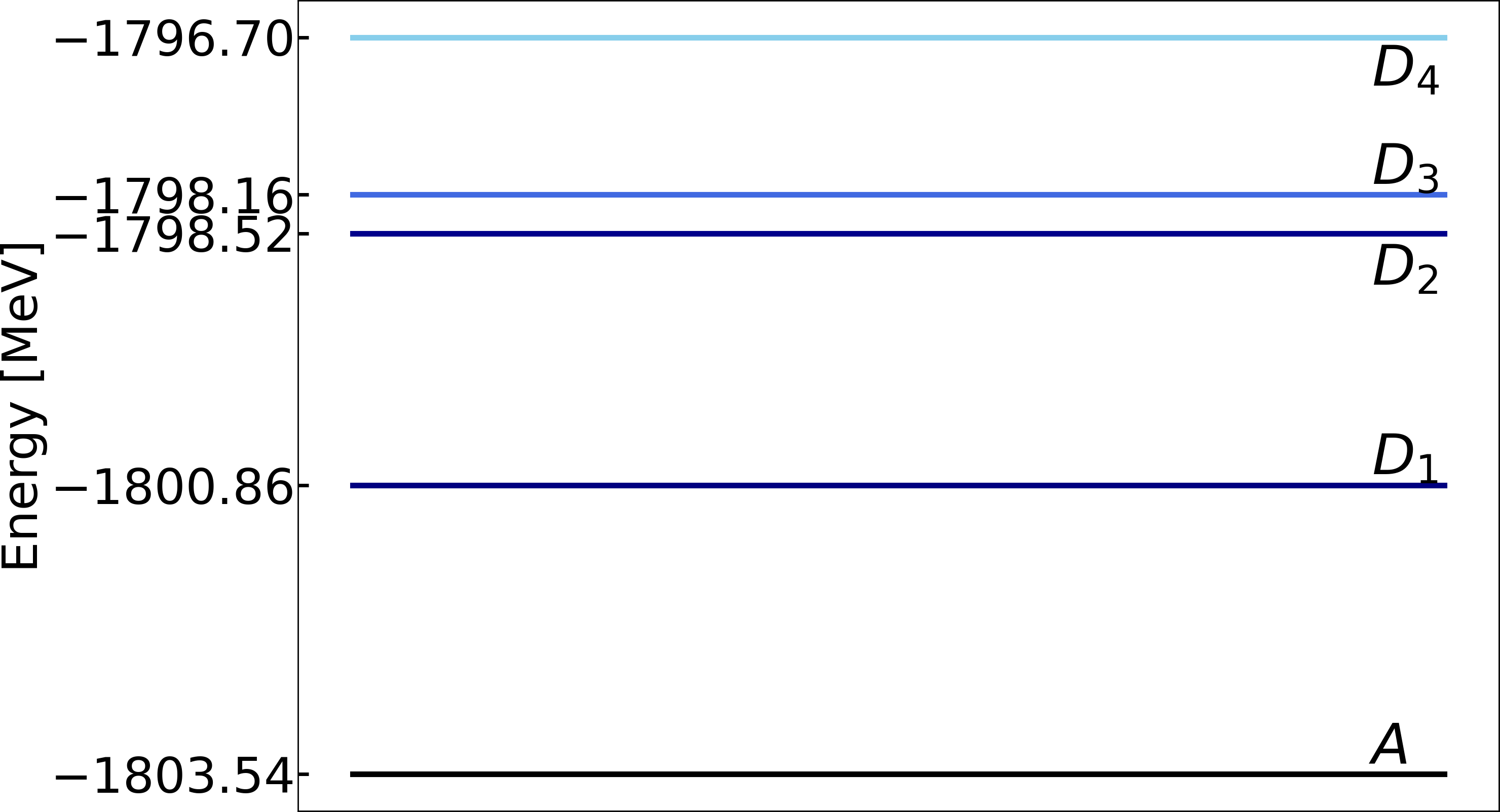}
\caption{Energy of four variational excited states (in blue) in $^{240}$Pu obtained with the \enquote{Deflation} method at its HFB ground state deformation (in black).}
\label{ctwo_66}
\end{figure}

An example of the \enquote{Deflation} procedure applied to generate several excited states in $^{240}$Pu above the ground-state HFB deformation is shown in FIG. \ref{ctwo_66}. In this calculation, orthogonality is imposed only within the full neutron isospin subspace, while all remaining degrees of freedom are left unconstrained. This procedure leads to a set of four variational excited states $\ket{D_i}$ constructed on top of the reference ground state $\ket{A}$.

As illustrated in FIG. \ref{ctwo_66}, the \enquote{Deflation} procedure generates a series of low-lying variational states with increasing excitation energy. The corresponding overlap matrix between the states is reported in TABLE \ref{ctwo_67}. The near-diagonal structure of this matrix confirms that the states are numerically almost orthogonal.

It should be emphasized that this example corresponds to a specific class of excitations and is not intended to represent the full low-energy spectrum of $^{240}$Pu. Nevertheless, it demonstrates the ability of the \enquote{Deflation} method to generate a hierarchy of constrained variational states in realistic nuclear systems.

The microscopic structure of such variational excitations is not straightforward and will be analyzed in detail in Section \ref{correlation} for the excited states generated by the \enquote{Continuous Deflation} method. This extension of the present approach, which constructs continuous excited states along an entire deformation path, is introduced below.

\begin{table}
\centering
\begin{tabular}{c |c c c c c}
\hline
& $\ket{A}$ & $\ket{D_1}$ & $\ket{D_2}$ & $\ket{D_3}$ & $\ket{D_4}$ \\
\hline
$\bra{A}$ & 1 &  4.1.10$^{-4}$ &  2.7.10$^{-28}$  & 8.0.10$^{-21}$ &  2.5.10$^{-5}$ \\
$\bra{D_1}$ & 4.1.10$^{-4}$ & 1 & 3.5.10$^{-31}$ & 7.7.10$^{-23}$ & 1.6.10$^{-5}$  \\
$\bra{D_2}$ & 2.7.10$^{-28}$ & 3.5.10$^{-31}$ & 1 &  1.4.10$^{-5}$  &7.9.10$^{-27}$  \\
$\bra{D_3}$ & 8.0.10$^{-21}$ & 7.7.10$^{-23}$ & 1.4.10$^{-5}$ & 1 & 9.4.10$^{-19}$ \\
$\bra{D_4}$ & 2.5.10$^{-5}$ & 1.6.10$^{-5}$ & 7.9.10$^{-27}$ & 9.4.10$^{-19}$ & 1 \\
\hline
\end{tabular}
\caption{Overlaps between the states created with the \enquote{Deflation} method in the $^{240}$Pu nucleus.}
\label{ctwo_67}
\end{table}

\subsection{\enquote{Continuous Deflation} method}  \label{ctwo_65}

\begin{figure}
\centering
\includegraphics[width=1.0\linewidth]{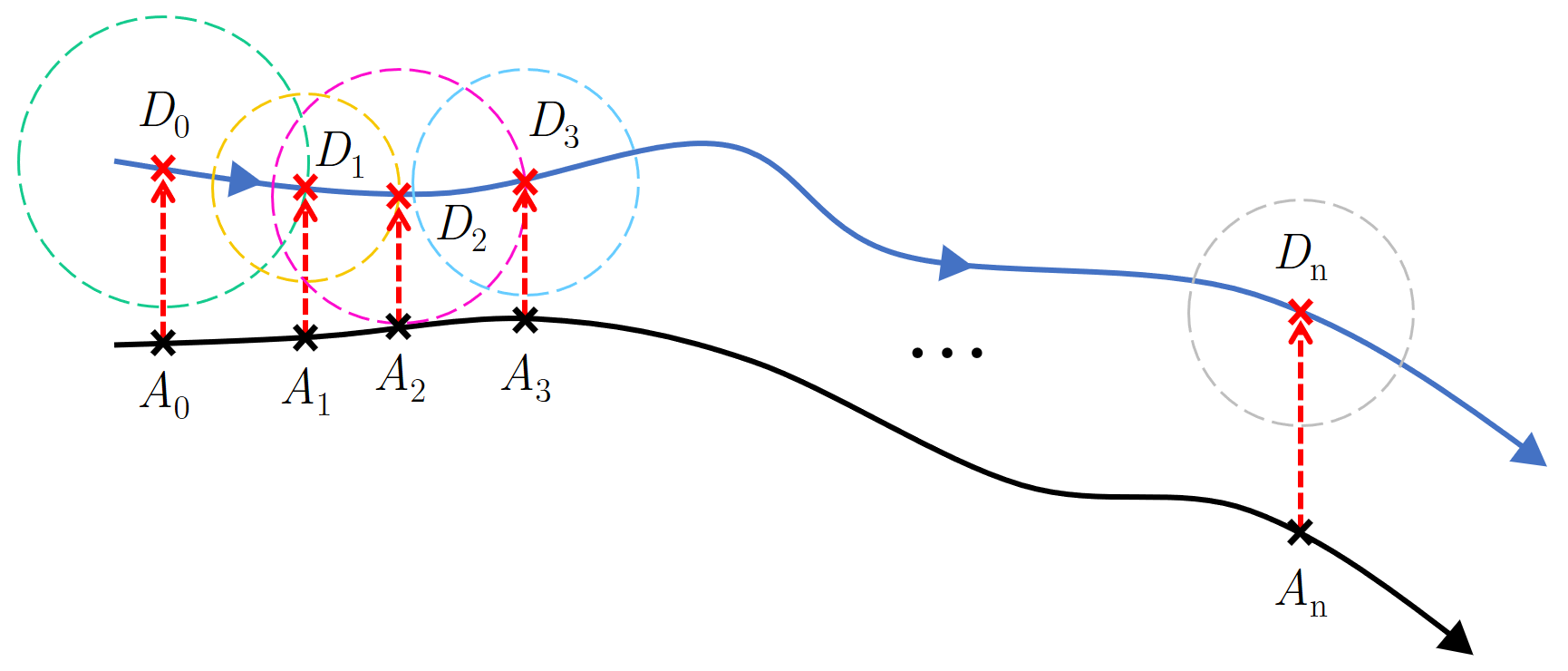}
\caption{Schematic representation of the \enquote{Continuous Deflation} method.}
\label{ctwo_59}
\end{figure}

The \enquote{Continuous Deflation} method constructs a sequence of excited states $\{ \ket{D_i} \}$ associated with a continuous adiabatic reference path $\{ \ket{A_i} \}$. In the way we have developed it in view of SCIM applications (specific choice), it enforces, for all $i$, a fixed overlap between consecutive excited states,
$\bra{D_i}\ket{D_{i+1}} = x_0$, together with a vanishing overlap with the adiabatic reference in a selected $(\tau^*, \Omega^*)$ subspace. Thus:
\begin{equation}
\bra{A^{\tau^*\Omega^*}_i}\ket{D^{\tau^*\Omega^*}_i} = 0.
\end{equation}
The parameter $x_0$ is chosen to match the overlap between consecutive adiabatic states (here, $x_0=0.995$), ensuring that both sets share the same collective coordinate, as required in the SCIM framework. In practice, the method is initialized from a seed state $\ket{D_0}$ constructed at the saddle point on the reference adiabatic state $\ket{A_0}$. This choice is motivated by the relevance of low-lying excitations in the descent from saddle to scission containing the pair-breaking phenomenon. The full procedure is illustrated in FIG. \ref{ctwo_59}.

\begin{figure}
\centering
\includegraphics[width=1.0\linewidth]{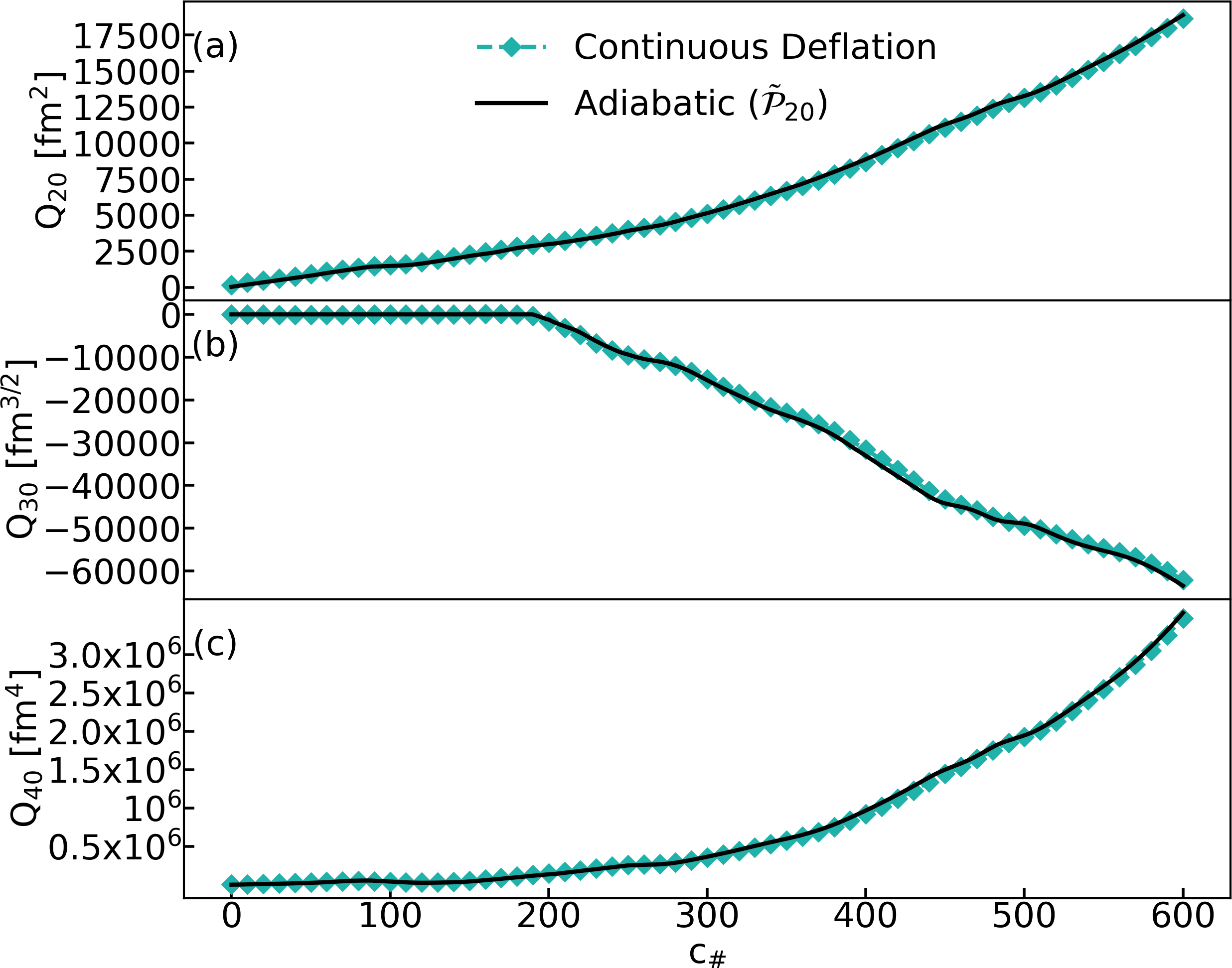}
\caption{Panel (a): Evolution of the quadrupole moment $Q_{20}$ of the $\mathcal{\tilde{P}}_{20}$ adiabatic set in $^{240}$Pu (in black) and the associated continuous neutron variational excitation one according to the new collective variable $c_\#$ defined in Ref. \cite{trilogy1}.  Panel (b): Same as panel (a) for the octupole moment $Q_{30}$. Panel (c): Same as panel (a) for the hexadecapole moment $Q_{40}$.}
\label{ctwo_85}
\end{figure}

A key difficulty is to ensure that the generated states remain intrinsic excitations of their reference configuration, i.e. that they preserve, to a large extent, the same underlying collective structure. Direct constraints on multipole moments were found to over-constrain the problem when combined with continuity and orthogonality requirements. We therefore adopt a mixed constraint strategy: orthogonality is imposed in a selected $(\tau^*, \Omega^*)$ subspace, while the overlap with the complementary isospin subspace is constrained to unity. This method indirectly constrains the shape of the variational state, leveraging the relatively uniform neutron-to-proton ratio in nuclei.
As illustrated in FIG. \ref{ctwo_85}, one sees that a resulting neutron excitation preserves a smooth evolution of the lowest multipole moments (quadrupole $Q_{20}$, octupole $Q_{30}$, and hexadecapole $Q_{40}$) consistent with the underlying adiabatic path. This indicates that the variational excited states follow the global deformation of the system.

When constructing variational excitations, mutual orthogonality would in principle be required between all states. A key advantage of the present scheme is that neutron and proton excitations are automatically orthogonal by construction. Within a given isospin subspace, states associated with different $\Omega$ blocks are found to remain nearly orthogonal even without additional constraints, with maximal overlaps below 0.033 for the ten excitations considered thereafter. This indicates that enforcing orthogonality in a single subspace induces a broadly distributed rearrangement of the mean field, rather than a selective mixing between specific channels. For these reasons, we do not impose additional mutual orthogonality constraints between the different excitations generated in this work. The main limitation of this choice is that it restricts the construction to at most one excitation per $(\tau, \Omega)$ subspace. Extensions of the algorithm are planed in future studies to extract several excitations by 
$\Omega$ blocks.

\begin{figure}
\centering
\includegraphics[width=1.0\linewidth]{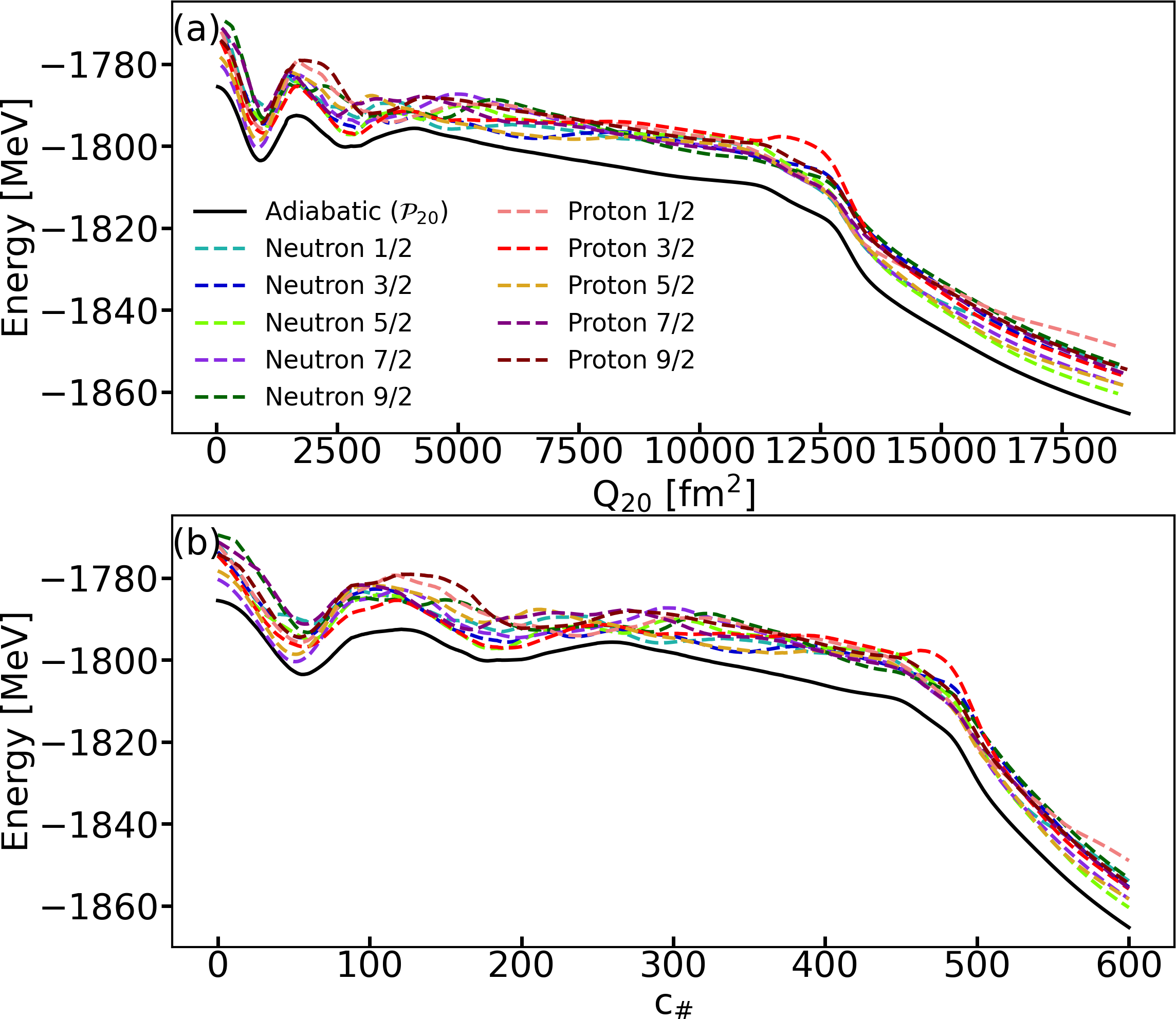}
\caption{Panel (a): Potential energy paths of ten variational excited states in $^{240}$Pu as a function of the quadrupole deformation $Q_{20}$. The $\mathcal{\tilde{P}}_{20}$ adiabatic surface is shown in black.
Panel (b): Same set of states represented in terms of the collective coordinate $c_\#$.}
\label{ctwo_61}
\end{figure}

In this work, we have generated up to ten variational excited sets for the $^{240}$Pu nucleus, spanning both isospins and $\Omega$ values from $1/2$ to $9/2$, with seeds initialized at the saddle point ($Q_{20} = 4230$ fm$^2$). FIG. \ref{ctwo_61} displays the corresponding potential energy paths, together with the adiabatic one (in black), as a function of the quadrupole deformation $Q_{20}$ in panel (a) and the collective coordinate $c_\#$ in panel (b).
The most striking feature in panel (a) is the convergence of the excited state paths near the scission region ($Q_{20} \approx 13000$ fm$^2$), followed by a subsequent splitting, which suggests an increased low-energy level density close to scission.
Panel (b) shows the same results in the $c_\#$ representation, which is more relevant for dynamical applications. Two main differences emerge: a broader first barrier and a shorter, steeper descent from saddle to scission.

\begin{figure}
\centering
\includegraphics[width=1.0\linewidth]{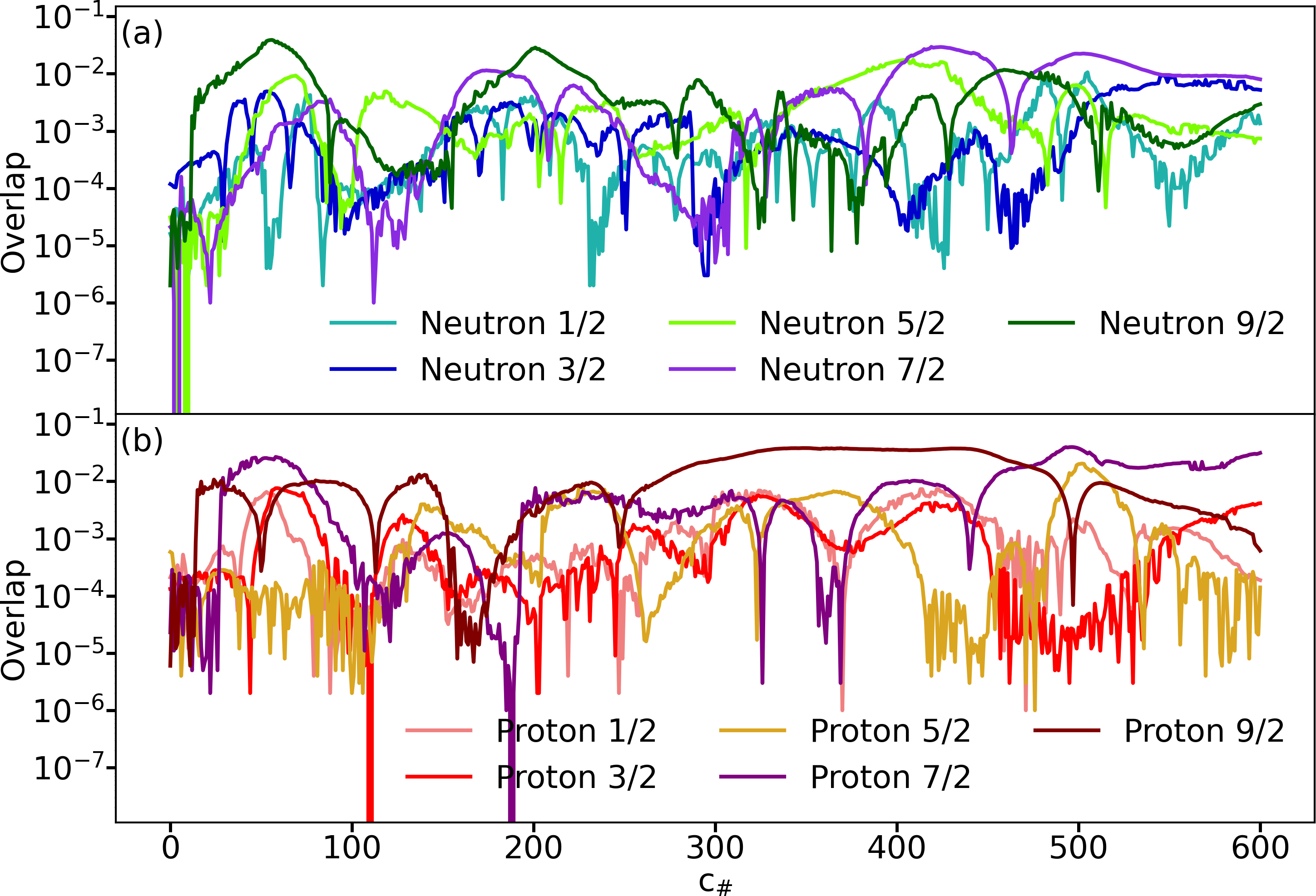}
\caption{Overlap between each variational excited state and its corresponding adiabatic state as a function of the collective variable $c_\#$. Panel (a): neutron excitations. Panel (b): proton excitations.}
\label{ctwo_90}
\end{figure}

Finally, we have examined the orthogonality between the variational excitations and their corresponding adiabatic states. In FIG. \ref{ctwo_90}, the overlap between each variational excited state and its associated adiabatic reference is shown as a function of $c_\#$. 
The overlap exhibits non-negligible variations along the excitation paths, with an average value of approximately $5 \times 10^{-4}$. For states associated with $\Omega = 1/2$, $3/2$, and $5/2$, the overlaps systematically remain below $10^{-2}$.
Assessing whether this level of orthogonality is satisfactory is not straightforward. However, such overlap values would correspond to a significant discontinuity if observed between neighboring states along an adiabatic potential energy surface. We therefore consider that the present construction provides a sufficiently clear separation between adiabatic and excited states, preventing ambiguities such as double counting in subsequent dynamical applications. 

From a numerical standpoint, the HO basis of the excited state at a given deformation is always taken to be the same as for the corresponding adiabatic state, which limits unnecessary numerical complications. However, when the single-particle bases of two consecutive adiabatic configurations $\ket{A_{i-1}}$ and $\ket{A_i}$ differ, additional numerical difficulties arise, which are explicitly accounted for in the implementation \cite{TPaul}.
Despite these technical difficulties, the method remains numerically robust. From a computational point of view, the construction of a full excited path containing more than 700 states for $^{240}$Pu requires approximately 10 hours for a 2×11 harmonic oscillator basis on a standard laptop with one core (no parallelization).

\section{Structure and regularity of the Continuous Deflation excitations along the $^{240}$Pu fission path}\label{correlation}

In this section, we analyze the content of the paths
we have built previously as excited states above the $^{240}$Pu adiabatic one. In particular, we study their
nature in terms of QP excitations relatively to the adiabatic path. Besides, we establish correlations between several indicators used in the analysis. Then, 
we evaluate the kernel regularity built from them. 

\subsection{Microscopic structure of the Continuous Deflation excitations}

\subsubsection{Analysis tools}

To characterize the microscopic structure of the variational excited states generated by the Continuous Deflation method, we introduce five complementary quantities probing different aspects of their QP content and excitation mechanism. The first quantity is the 2QP content:
\begin{eqnarray}\label{ctwo_78}
\sigma^{(2)}
=
\sum_{ij}
\left|
\bra{\Phi^*}
\xi_i^+
\bar{\xi}_j^+
\ket{\Phi}
\right|^2,
\end{eqnarray}
where $\ket{\Phi}$ and $\ket{\Phi^*}$ denote the reference adiabatic state and its associated variational excited state, respectively. This quantity measures the fraction of a variational excited state that can be described as a superposition of 2QP excitations built on the HFB reference state $\ket{\Phi}$.

Higher-order correlations are characterized through the 4QP content in the following analysis:
\begin{align}\label{ctwo_84}
&\sigma^{(4)}
= \frac{1}{2}
\sum_{\alpha\beta}
\sum_{ij,\;
(\Omega_{\alpha\beta},\tau_{\alpha\beta})
\neq
(\Omega_{ij},\tau_{ij})}
\left|
\bra{\Phi^*}
\xi_\alpha^+
\bar{\xi}_\beta^+
\xi_i^+
\bar{\xi}_j^+
\ket{\Phi}
\right|^2
\nonumber\\
& \; \; \;+
\frac{1}{4}
\sum_{\alpha\beta}
\sum_{ij,\;
(\Omega_{\alpha\beta},\tau_{\alpha\beta})
=
(\Omega_{ij},\tau_{ij})}
\left|
\bra{\Phi^*}
\xi_\alpha^+
\bar{\xi}_\beta^+
\xi_i^+
\bar{\xi}_j^+
\ket{\Phi}
\right|^2,
\end{align}
which probes the leading corrections beyond the 2QP picture. Throughout this section, we also consider the cumulative QP content $\sigma^{{tot}}$ defined by: 
\begin{align}
   \sigma^{{tot}}
=
\sigma^{(2)}
+
\sigma^{(4)},
\end{align}
which estimates the total fraction of the wave function described by the lowest QP orders.

To quantify the structural rearrangement induced by the orthogonality constraint, we introduce the purity indicator:
\begin{eqnarray}
O^r
=
\prod_{(\tau,\Omega)\neq(\tau^*,\Omega^*)}
\bra{\Phi^{*\tau\Omega}}
\ket{\Phi^{\tau\Omega}}.
\end{eqnarray}
This last quantity measures how strongly an excitation modifies the unconstrained $(\tau,\Omega)$ subspaces. Values close to unity therefore correspond to nearly pure excitations, whereas smaller values indicate stronger rearrangements of the unconstrained subspaces.

Finally, we monitor the excitation energy:
\begin{eqnarray}
\Delta E^*=E^*-E,
\end{eqnarray}
where $E^*$ and $E$ represents the total energy of the excited and adiabatic states, respectively.
This last quantity provides a complementary energetic characterization of the variational excited states.

These four quantities constitute the main analysis tools used throughout the remainder of this section. We first examine their evolution along each excited path before investigating the statistical correlations that emerge from the complete set of variational excitations.

\subsubsection{Microscopic composition of the variational excited paths}

\begin{figure}
\centering
\includegraphics[width=1.0\linewidth]{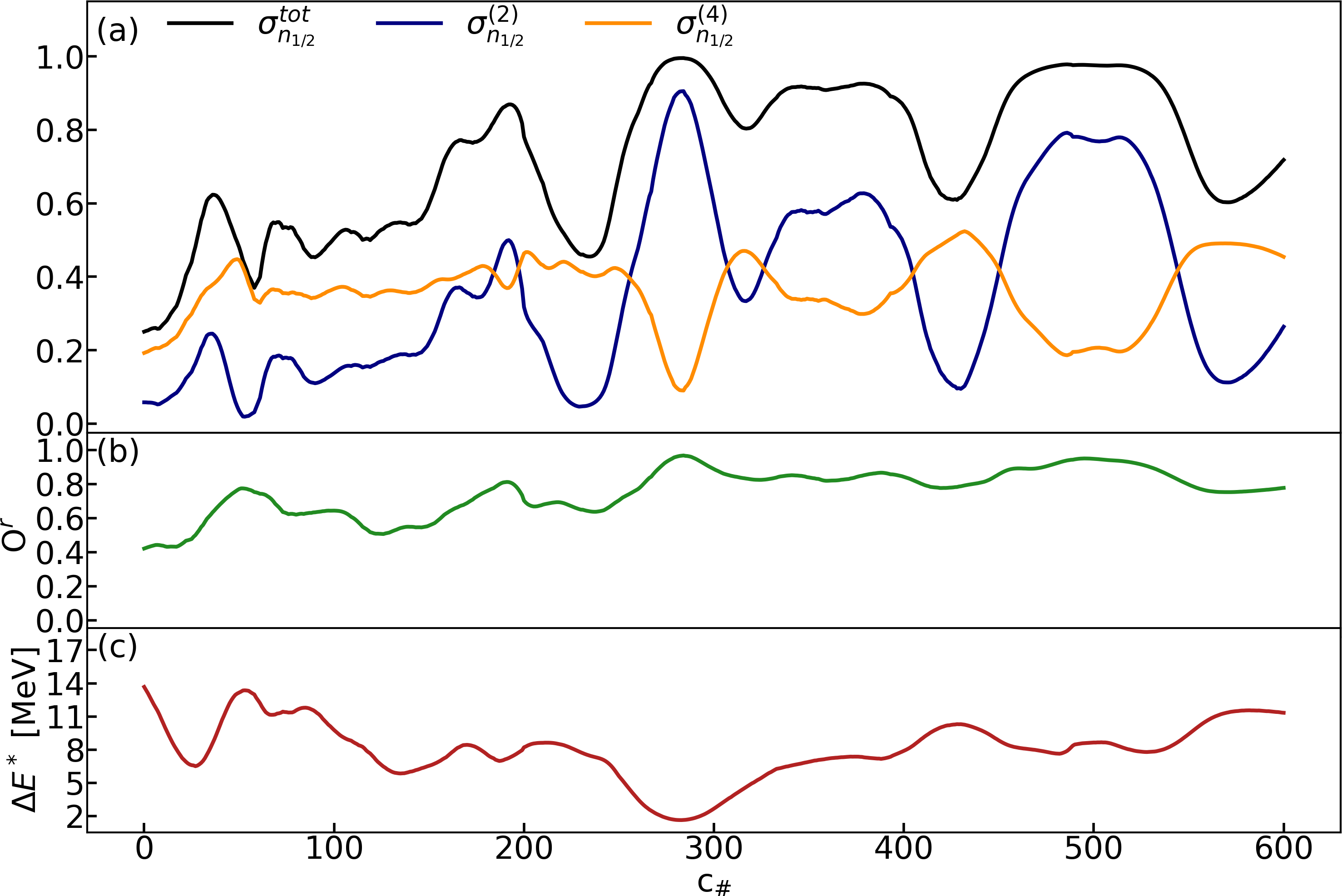}
\caption{Panel (a): Evolution of $\sigma^{tot}$ (in black), $\sigma^{(2)}$ (in blue) and $\sigma^{(4)}$ (in orange) in the neutron variational excitation associated with $\Omega = 1/2$ as a function of $c_\#$. Panel (b): Evolution of $O^r$ (in green). Panel (c): Evolution of $\Delta E^*$ (in brown) expressed in MeV.}
\label{ctwo_91}
\end{figure}

\begin{figure}
\centering
\includegraphics[width=1.0\linewidth]{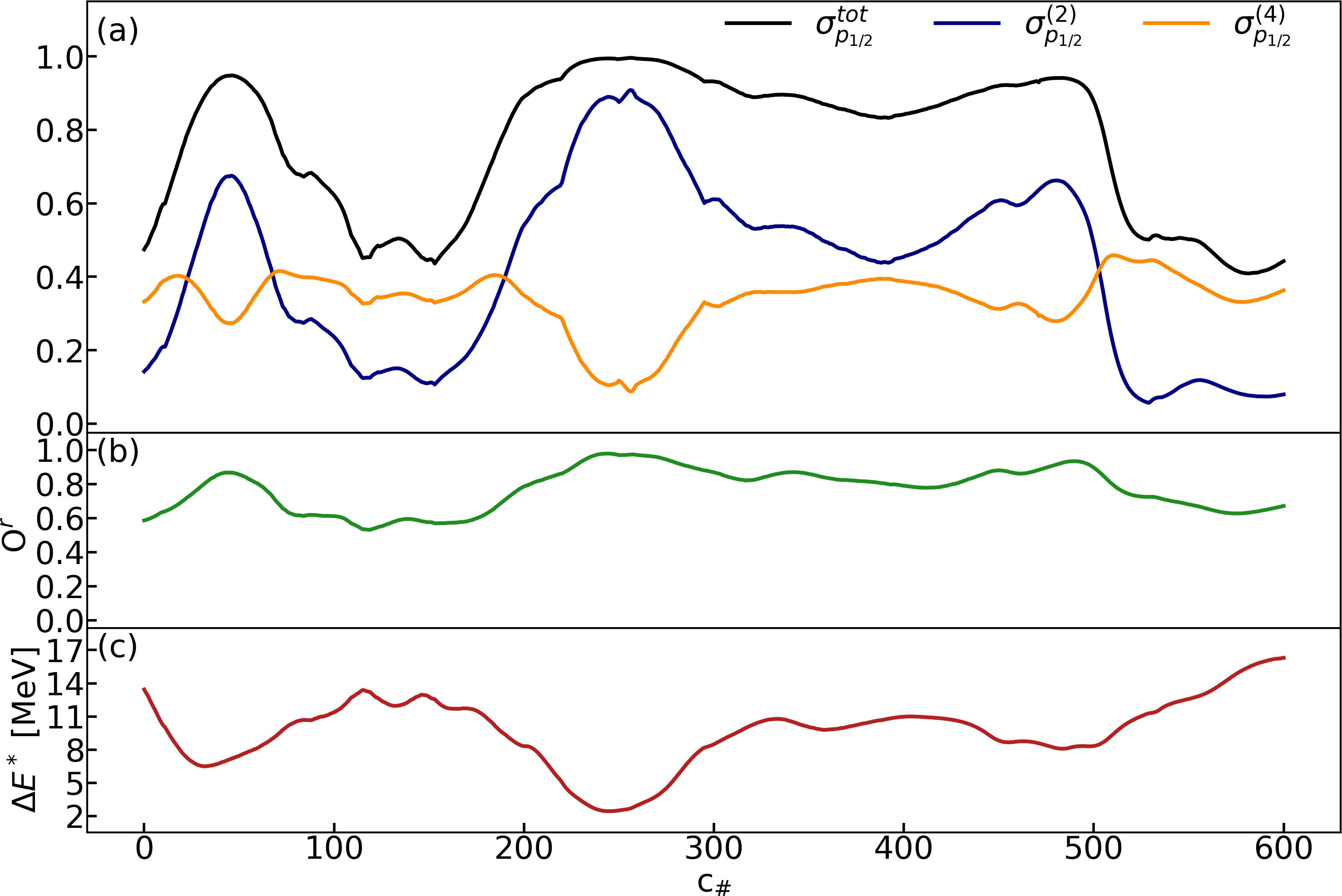}
\caption{Same as FIG.\ref{ctwo_91} but for the proton variational excitation associated with $\Omega = 1/2$.}
\label{ctwo_114}
\end{figure}

\begin{figure}
\centering
\includegraphics[width=1.0\linewidth]{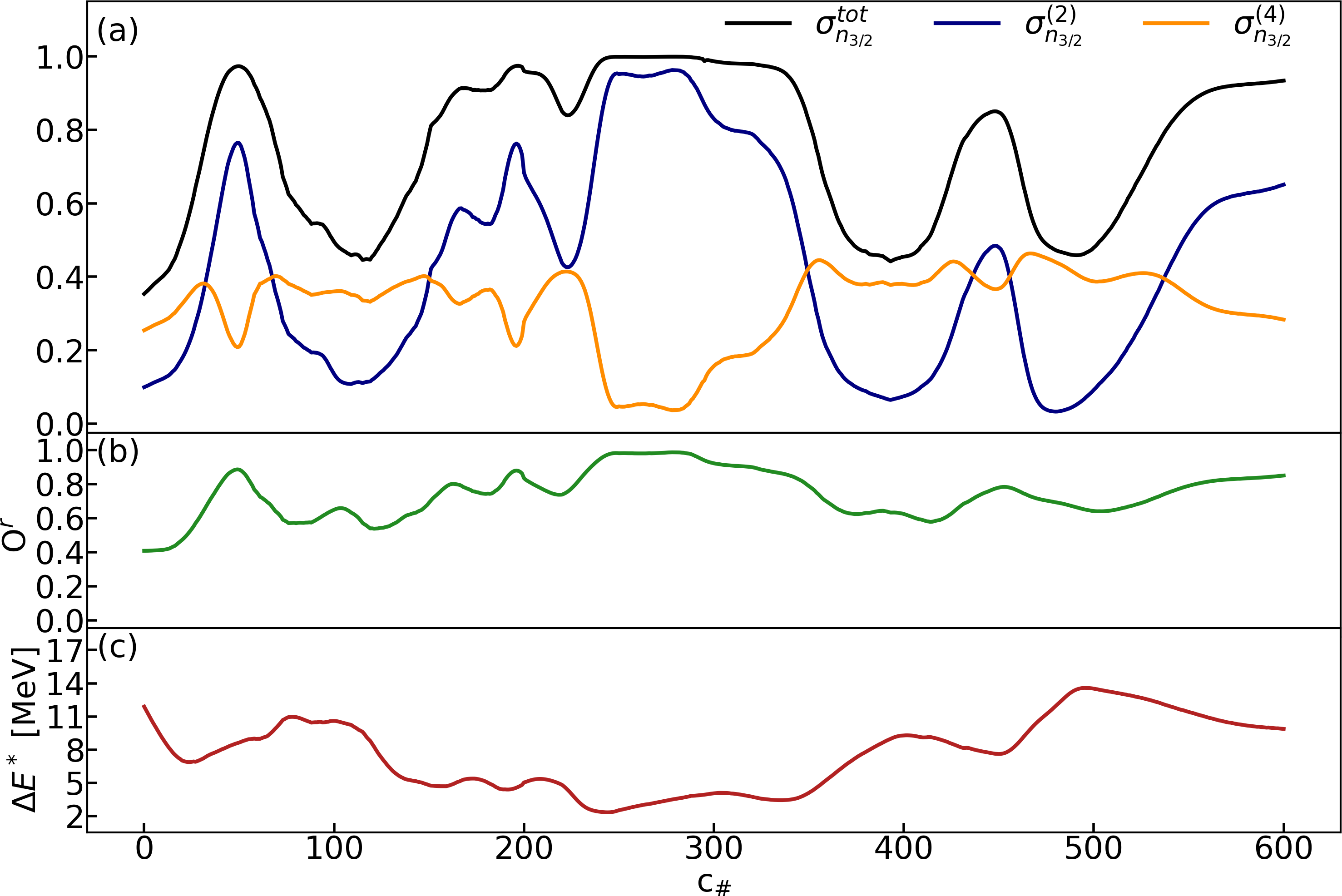}
\caption{Same as FIG.\ref{ctwo_91} but for the neutron variational excitation associated with $\Omega = 3/2$.}
\label{ctwo_106}
\end{figure}

\begin{figure}
\centering
\includegraphics[width=1.0\linewidth]{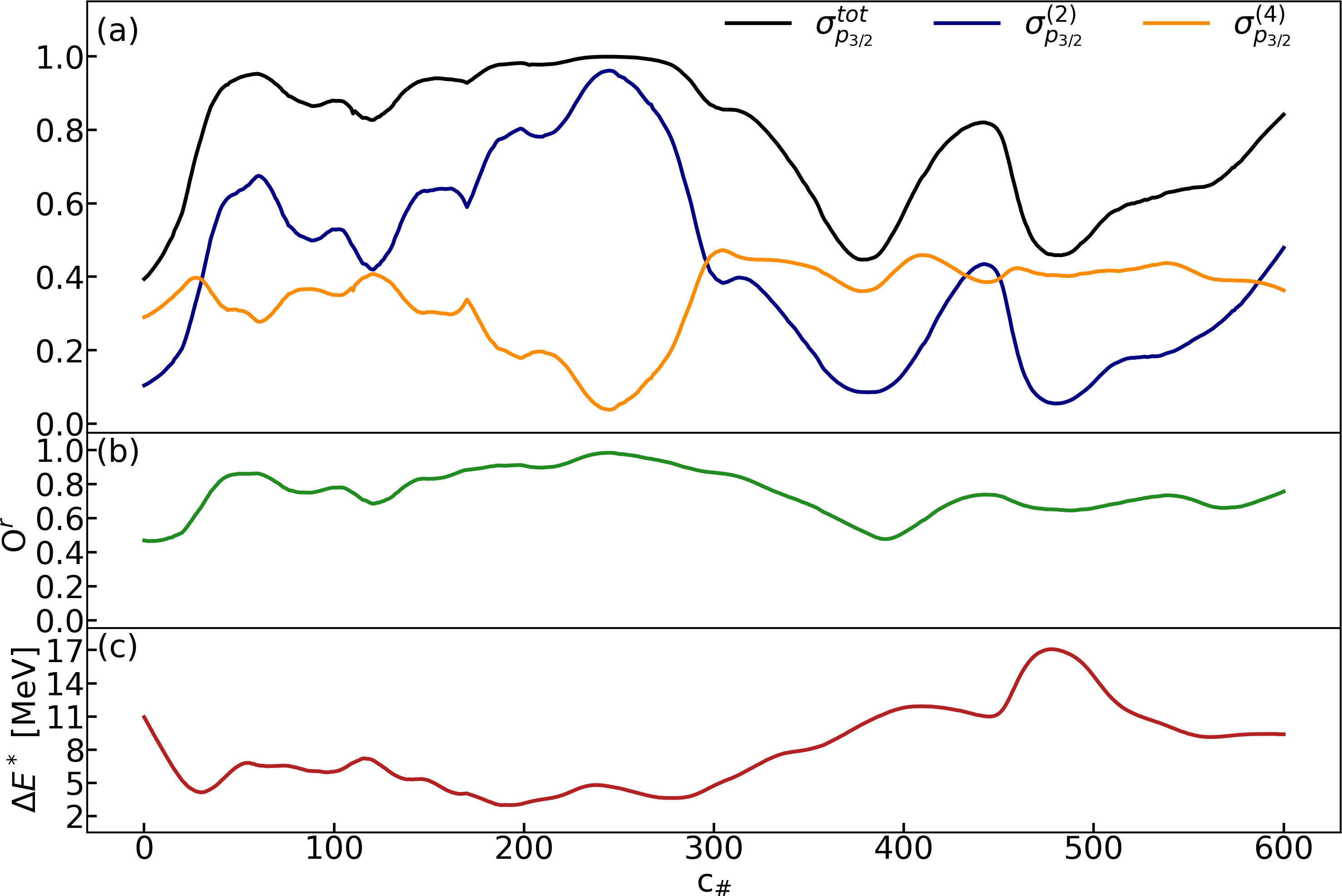}
\caption{Same as FIG.\ref{ctwo_91} but for the proton variational excitation associated with $\Omega = 3/2$.}
\label{ctwo_111}
\end{figure}

\begin{figure}
\centering
\includegraphics[width=1.0\linewidth]{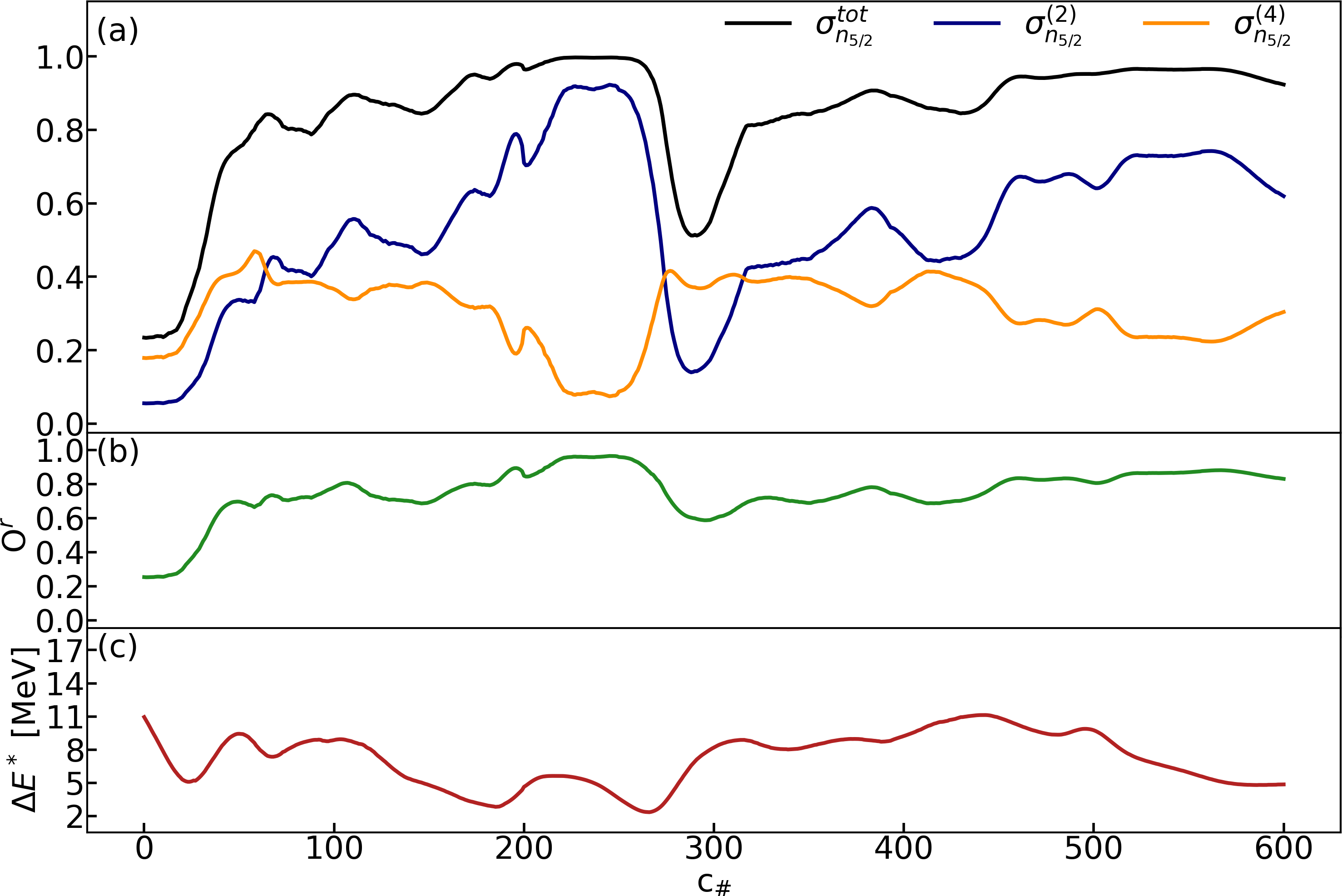}
\caption{Same as FIG.\ref{ctwo_91} but for the neutron variational excitation associated with $\Omega = 5/2$.}
\label{ctwo_107}
\end{figure}

\begin{figure}
\centering
\includegraphics[width=1.0\linewidth]{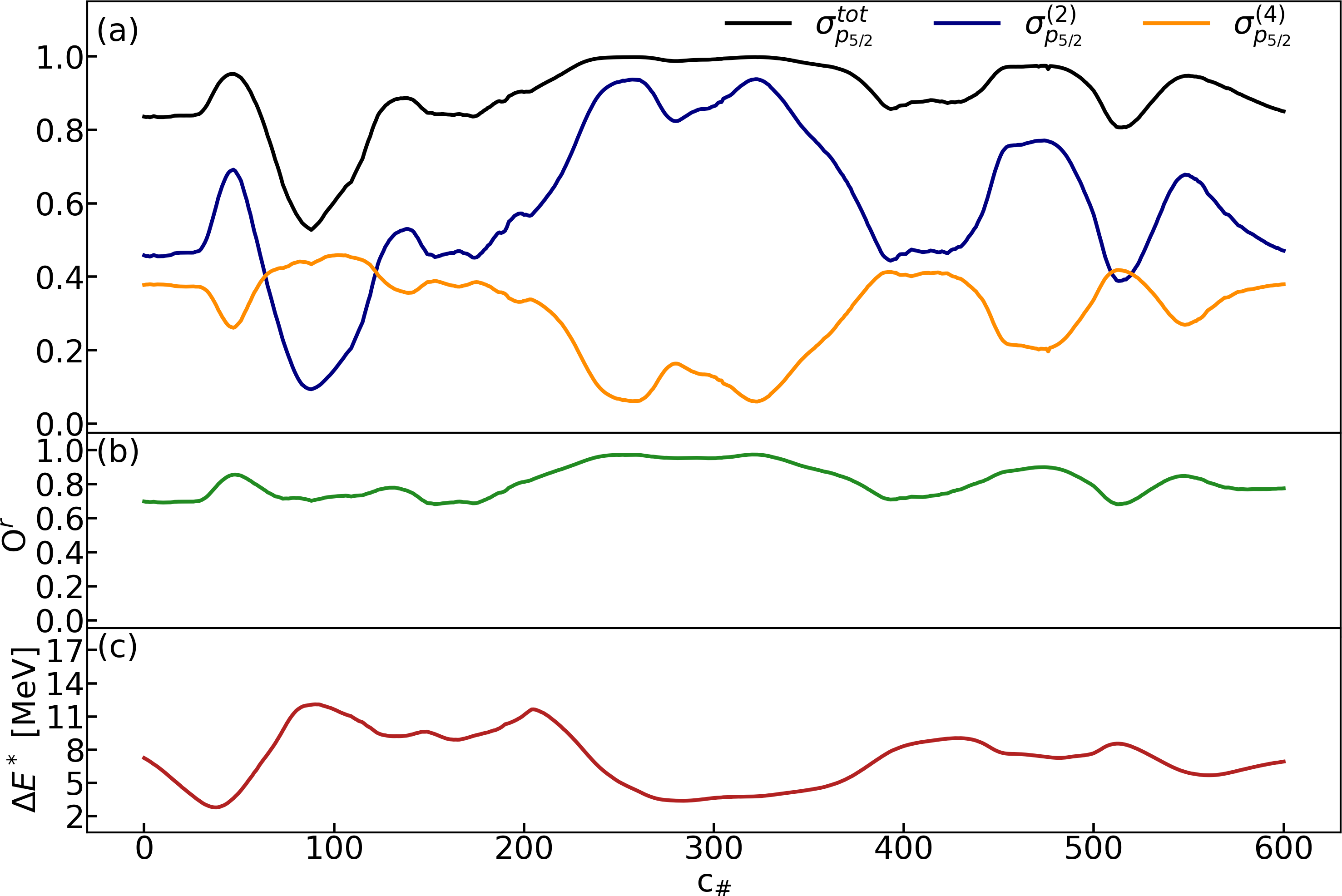}
\caption{Same as FIG.\ref{ctwo_91} but for the proton variational excitation associated with $\Omega = 5/2$.}
\label{ctwo_112}
\end{figure}

\begin{figure}
\centering
\includegraphics[width=1.0\linewidth]{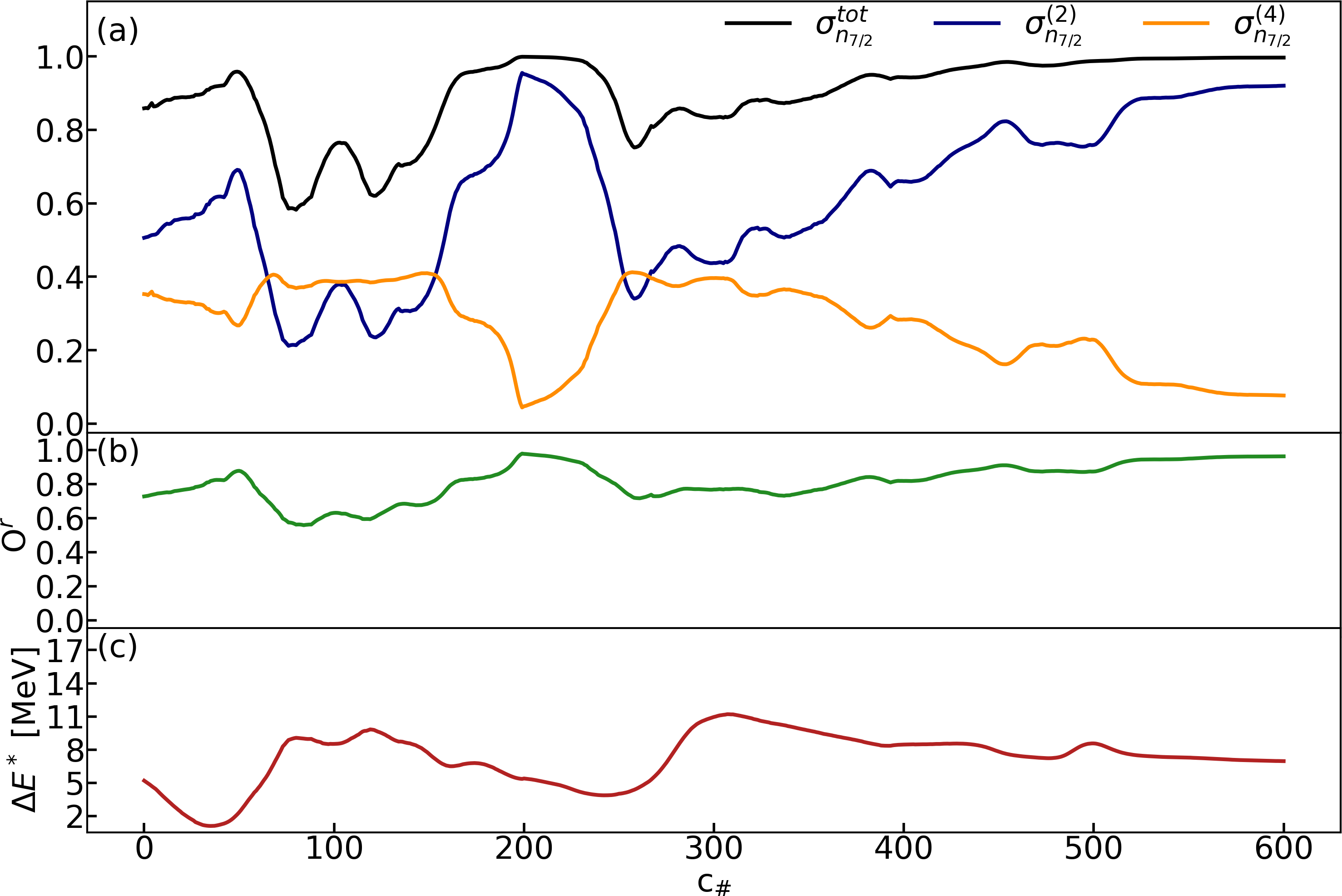}
\caption{Same as FIG.\ref{ctwo_91} but for the neutron variational excitation associated with $\Omega = 7/2$.}
\label{ctwo_108}
\end{figure}

\begin{figure}
\centering
\includegraphics[width=1.0\linewidth]{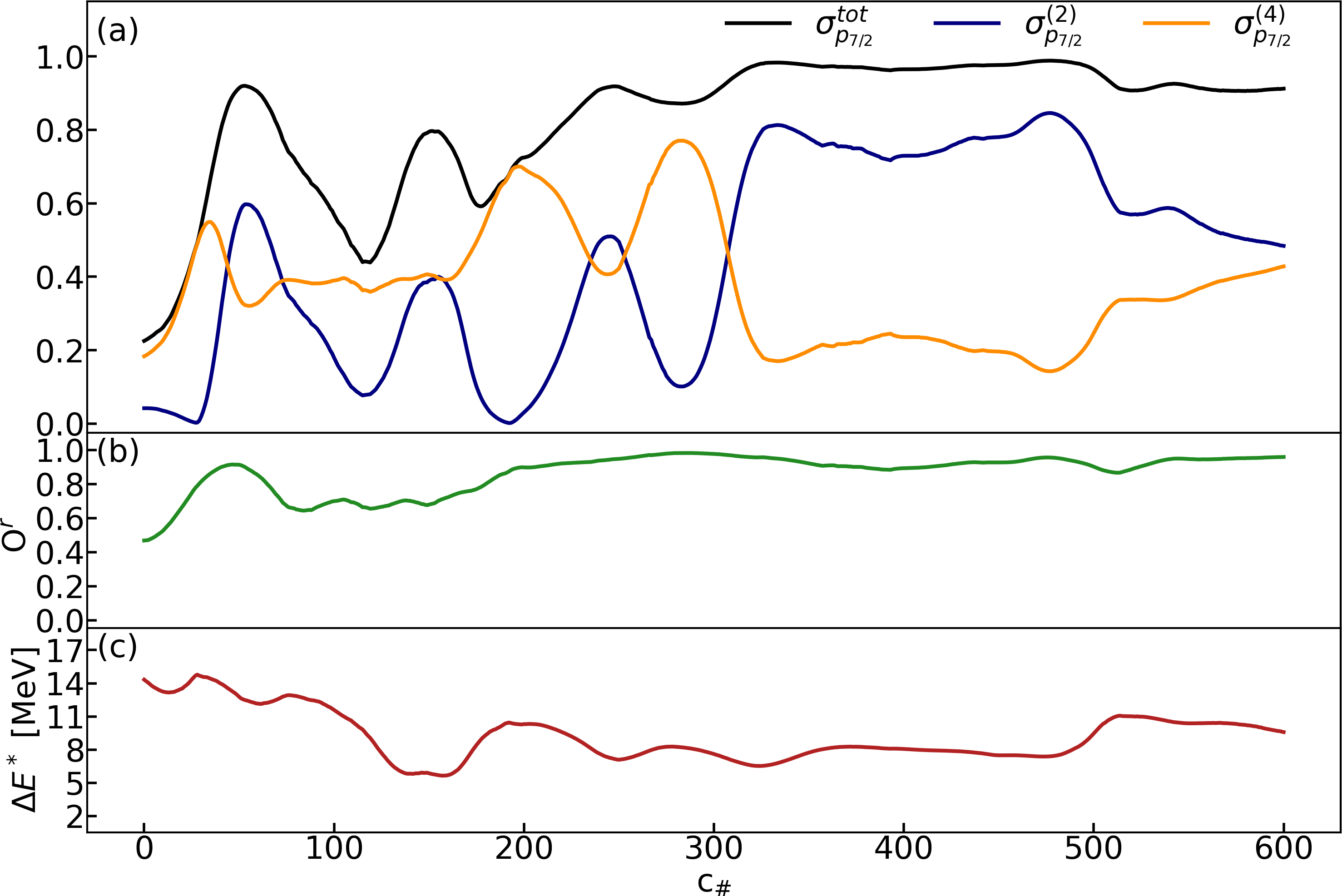}
\caption{Same as FIG.\ref{ctwo_91} but for the proton variational excitation associated with $\Omega = 7/2$.}
\label{ctwo_113}
\end{figure}

\begin{figure}
\centering
\includegraphics[width=1.0\linewidth]{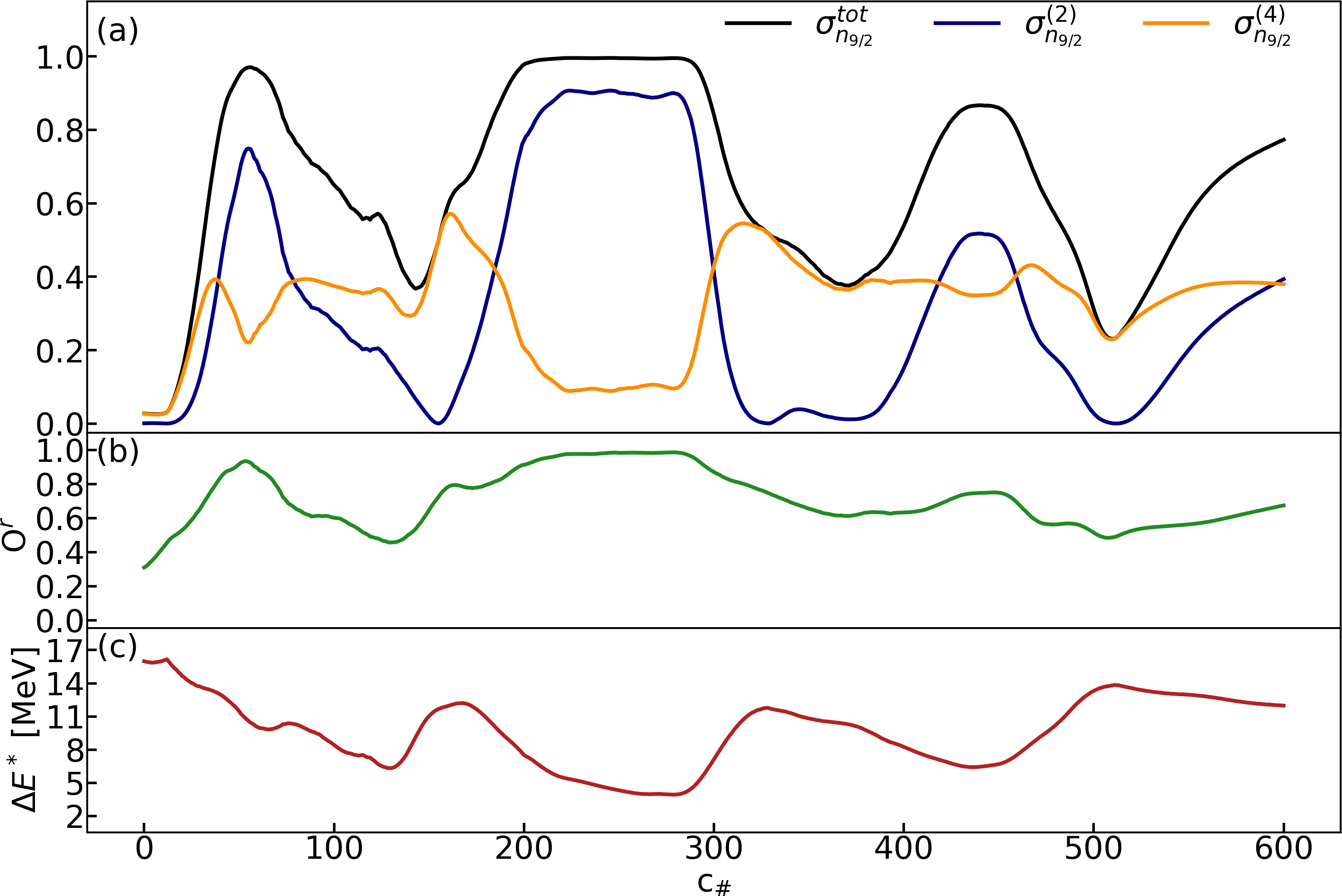}
\caption{Same as FIG.\ref{ctwo_91} but for the neutron variational excitation associated with $\Omega = 9/2$.}
\label{ctwo_109}
\end{figure}

\begin{figure}
\centering
\includegraphics[width=1.0\linewidth]{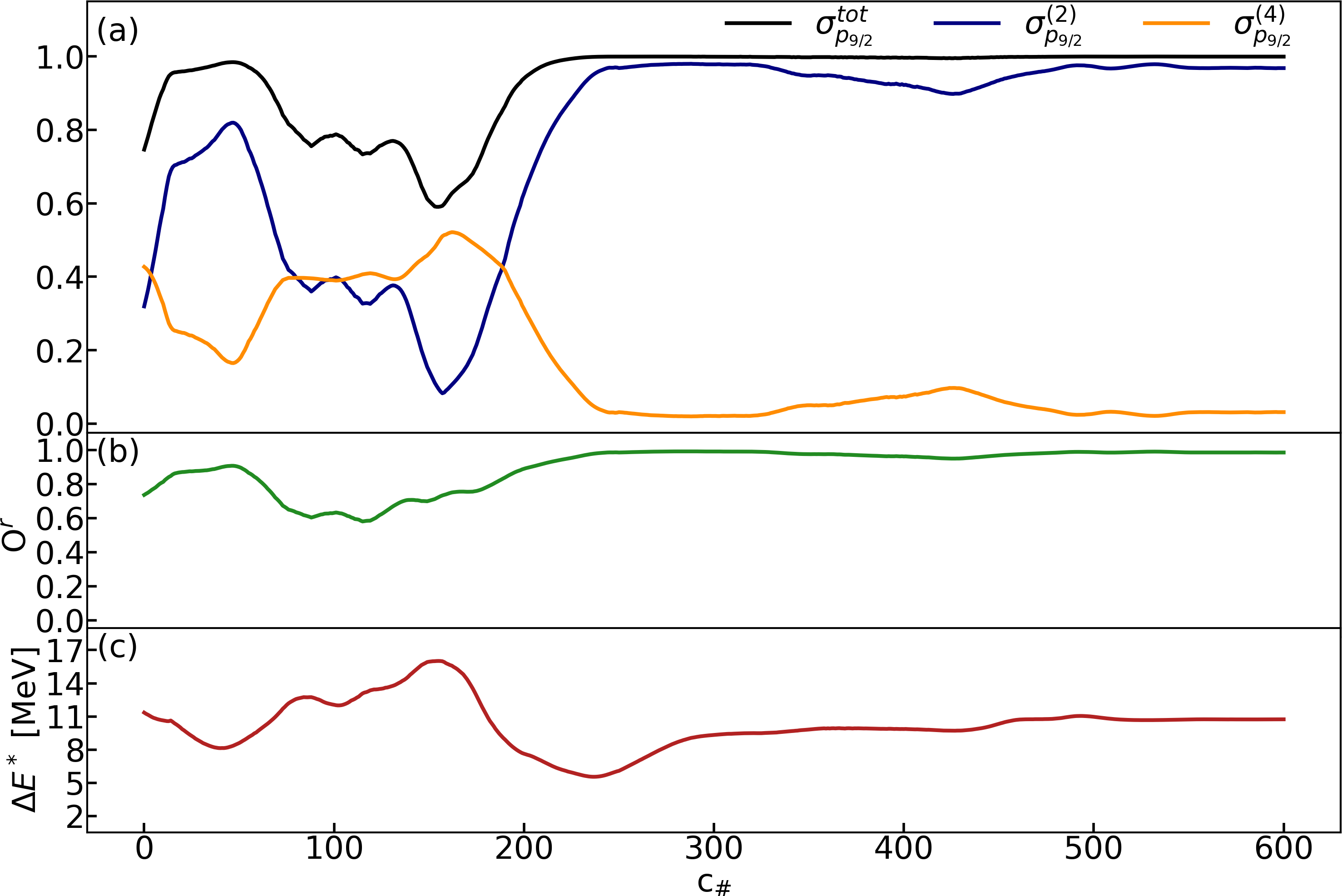}
\caption{Same as FIG.\ref{ctwo_91} but for the proton variational excitation associated with $\Omega = 9/2$.}
\label{ctwo_92}
\end{figure}

FIGs. \ref{ctwo_91} to \ref{ctwo_92} summarize the microscopic composition of the ten variational excited paths generated with the \enquote{Continuous Deflation} method. For each excitation, panel (a) displays the evolution of $\sigma^{(2)}$ (in blue), $\sigma^{(4)}$ (in orange), and $\sigma^{\mathrm{tot}}$ (in black), while panel (b) shows the purity indicator $O^r$, and panel (c) reports the excitation energy $\Delta E^*$ (in MeV) as a function of the collective coordinate $c_\#$.

Several general features emerge from these results. First, the combined contribution of the 2QP and 4QP components remains significant over most of the deformation path. For the majority of the generated excitations, $\sigma^{\mathrm{tot}}$ exceeds 50\%, indicating that the microscopic structure is largely captured by the first two QP orders.

Second, the relative importance of the 2QP and 4QP components evolves continuously along the fission path. Although the 2QP contribution dominates in many cases, several regions exhibit a comparable or even larger 4QP content, illustrating the progressive development of higher-order correlations as deformation increases. 

The evolution of the purity indicator $O^r$ broadly follows that of $\sigma^{{tot}}$, highlighting that the most weakly mixed excitations are also those best described within the truncated QP expansion. 

Finally, the excitation energies span a broad interval, typically between approximately 2 and 17 MeV, revealing a large diversity in the energetic properties of the generated excited paths.

Another remarkable feature concerns the behavior of the microscopic composition across the scission region ($c_\# \simeq 495$). Some excited paths preserve an almost unchanged QP composition through scission (see FIGs. \ref{ctwo_107}, \ref{ctwo_108}, \ref{ctwo_113}, and \ref{ctwo_92}), whereas others undergo pronounced structural rearrangements (see FIGs. \ref{ctwo_91}, \ref{ctwo_114}, \ref{ctwo_106}, \ref{ctwo_111}, and \ref{ctwo_109}). This observation suggests the existence of two classes of variational excitations. The first class would correspond to collective excitations involving both nascent fragments and would therefore be strongly affected by the rupture of the neck. The second class would instead correspond to excitations predominantly localized on one pre-fragment, whose microscopic structure remains largely preserved through scission. This interpretation will be further investigated in Section \ref{scissionarea}.

\subsubsection{Correlations between the microscopic indicators}

The evolution along the different excited paths already suggests that the microscopic quantities introduced above are not independent. To identify the dominant trends quantitatively, we now analyze the complete set of data generated with the Continuous Deflation method. 

\begin{figure}
\centering
\includegraphics[width=1.0\linewidth]{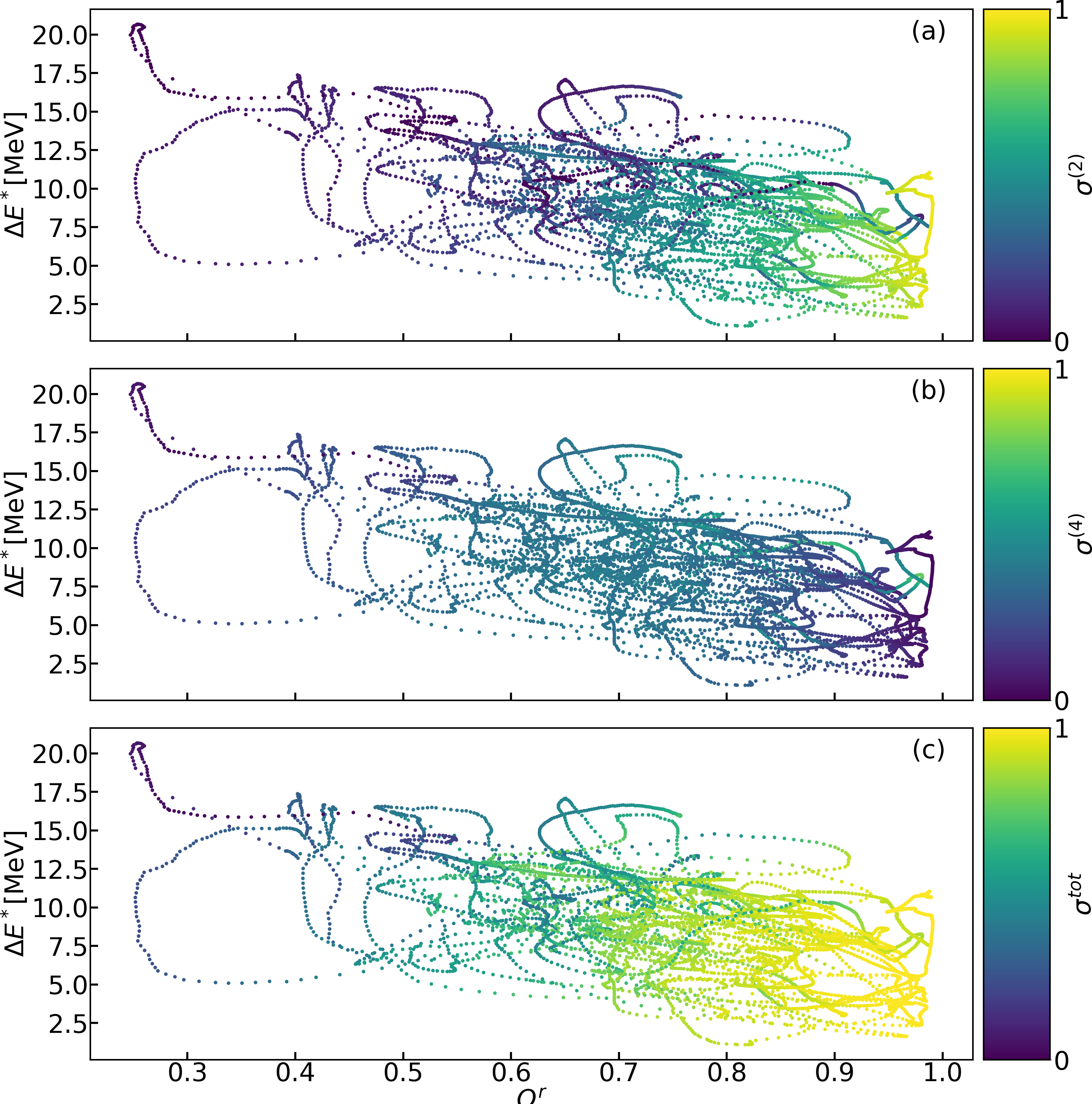}
\caption{Panel (a): Evolution of $\sigma^{(2)}$ with respect to both $O^r$ and the excitation energy $\Delta E^*$. Panel (b): Same as panel (a) but for $\sigma^{(4)}$. Panel (c): Same as panel (a) but for $\sigma^{tot}$.}
\label{ctwo_153}
\end{figure}

In FIG. \ref{ctwo_153}, we have gathered all calculated configurations and compare the evolution of the excitation energy $\Delta E^*$ according to the purity indicator $O^r$, and the QP contents $\sigma^{(2)}$ (panel (a)), $\sigma^{(4)}$ (panel (b)) and $\sigma^{\rm tot}$ (panel (c)).
The clearest correlation concerns the purity indicator $O^r$ and the total low-order QP content $\sigma^{\mathrm{tot}}$. A clear positive correlation is observed: states with large values of $O^r$ are also those exhibiting the largest values of $\sigma^{\mathrm{tot}}$. This confirms that excitations producing only limited rearrangements of the unconstrained subspaces are generally well described by low-order QP configurations. A similar trend is found between $O^r$ and $\sigma^{(2)}$, whereas intermediate values of $O^r$ appear to correlate more strongly with increased $\sigma^{(4)}$, indicating a progressive shift toward more complex configurations as the purity is partially reduced.

The excitation energy exhibits only moderate correlations with the QP content. While low-energy excitations are often associated with relatively large values of $\sigma^{(2)}$, intermediate excitation energies tend to favor larger values of $\sigma^{(4)}$. Nevertheless, a substantial dispersion is observed, indicating that the microscopic structure of the variational excitations cannot be inferred from their excitation energies alone. A similarly moderate correlation is found between the excitation energy and $O^r$, with the lowest-energy excitations tending to be the most pure.

\begin{figure}
\centering
\includegraphics[width=1.0\linewidth]{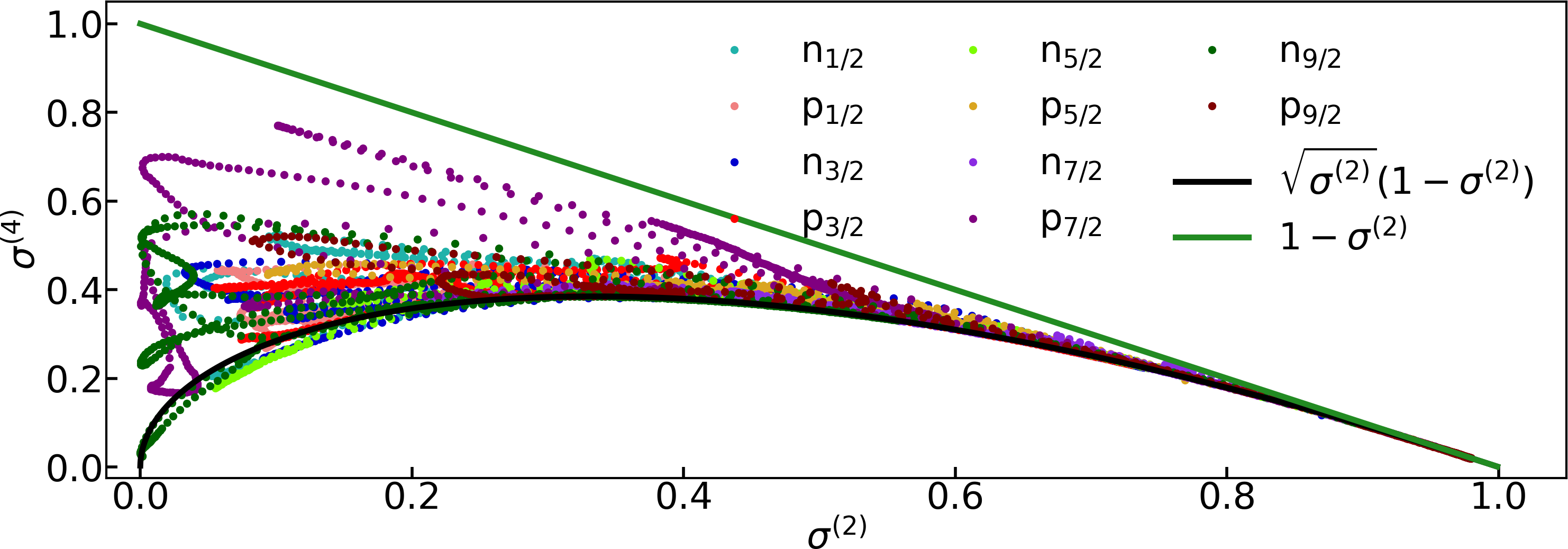}
\caption{Correlations between $\sigma^{(4)}$ and $\sigma^{(2)}$.}
\label{ctwo_142}
\end{figure}

To further examine the respective roles of the leading QP components, FIG. \ref{ctwo_142} shows the correlation between $\sigma^{(2)}$ and $\sigma^{(4)}$.

No strict one-to-one correlation is observed between the individual 2QP and 4QP contributions. Instead, these quantities tend to compensate one another, reflecting a gradual redistribution of the wave-function content between different QP orders along the deformation path. Interestingly, all calculated states are found to satisfy the empirical relation:
\begin{equation}
  \sigma^{(4)} \gtrsim \sqrt{\sigma^{(2)}}\left(1-\sigma^{(2)}\right) 
\end{equation}
which appears to define a lower boundary for the distribution of QP content. Although this relation has emerged consistently from the present calculations, we are not yet able to provide a microscopic interpretation of its origin. Its apparent universality nevertheless suggests that it reflects a deeper structural property of the variational excited states generated by the \enquote{Continuous Deflation} method.

\subsection{Kernel regularity of the Continuous Deflation excitations}

\subsubsection{Overlap kernels}

\begin{figure}
\centering
\includegraphics[width=1.0\linewidth]{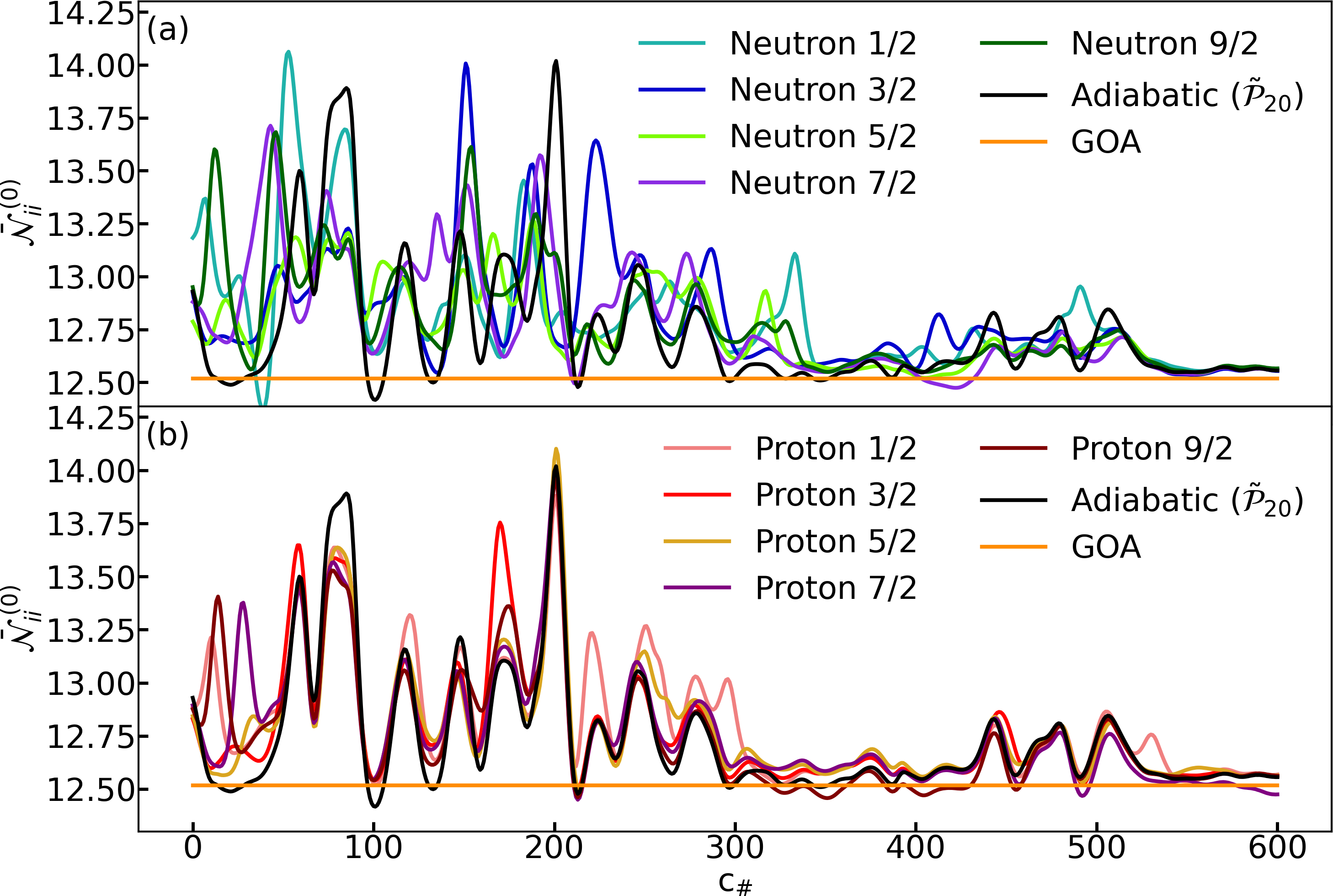}
\caption{Panel (a): Diagonal neutron overlap kernel moments of order zero $\mathcal{\bar N}_{ii}^{(0)}$ for all variational excitations as a function of $c_\#$ in $^{240}$Pu. Panel (b): Same as panel (a) but for protons. Results for the $ \mathcal{\tilde P}_{20}$ adiabatic set are also shown (in black), together with the corresponding GOA approximation (in orange).}
\label{ctwo_63}
\end{figure}

A key requirement for the SCIM framework to be effective is the regularity of the underlying overlap kernels (see Ref.\cite{trilogy1}, Eq. (14)). In this respect, we first examine the zero-order diagonal overlap kernel moments for all variational excitations, together with those of the associated adiabatic set, as a function of the collective coordinate $c_\#$:
\begin{eqnarray}
\mathcal{\bar N}_{00}^{(0)}(c_{\#}) = \int ds \bra{\Phi_0(c_{\#}-s)}\ket{\Phi_0(c_{\#}+s)}.
\end{eqnarray}
Results are displayed in FIG. \ref{ctwo_63}. The corresponding Gaussian Overlap Approximation (GOA) results are also shown for comparison. Panel (a) displays the neutron excitations, while panel (b) corresponds to protons.

We observe that the variational excitations exhibit a level of regularity in their overlap kernels that is similar to the one of the adiabatic reference set. It will be shown in the third article of the trilogy that the remaining variations are too large to allow for a well-defined extraction of the dynamical quantities $V_{\mathrm{SCIM}}$, $D_{\mathrm{SCIM}}$, and $B_{\mathrm{SCIM}}$, a limitation that is addressed therein through the application of a Savitzky--Golay low-pass filtering procedure~\cite{SGDif}.

Regarding non-diagonal overlap kernels, no well-established benchmark exists to assess the adequacy of their regularity. However, their zeroth- and second-order moments are typically one to four orders of magnitude smaller than those of the diagonal kernels. This hierarchy suggests that their contributions to the overall dynamical input is sub-leading. In practice, no numerical instabilities or pathological behaviors have been observed in the dynamical evolution associated with these non-diagonal terms.

\subsubsection{Hamiltonian kernels}

To investigate the regularity properties of the Hamiltonian kernels, we examine whether the adiabatic-level approximation relating Hamiltonian and overlap kernels remains valid for the variational excitations. This choice is motivated by the fact that the relevant property for dynamical applications is ultimately the regularity of the ratio between the Hamiltonian and overlap kernels. We therefore quantify deviations from this approximation through the relative deviation:
\begin{eqnarray}
\Delta \hat H_{ii}(\bar q-s,\bar q+s) = 100 \times \qquad \qquad \qquad \qquad \\
\left|\frac{\bra{\Phi_{i}(\bar q-s)}\hat H \ket{\Phi_i(\bar q+s)} - \bra{\Phi_i(\bar q-s)}\ket{\Phi_i(\bar q +s)}E_i(\bar q)}{\bra{\Phi_{i}(\bar q -s)}\hat H \ket{\Phi_i(\bar q +s)}}\right|.
\nonumber
\end{eqnarray}
\begin{figure}
\centering
\includegraphics[width=1.0\linewidth]{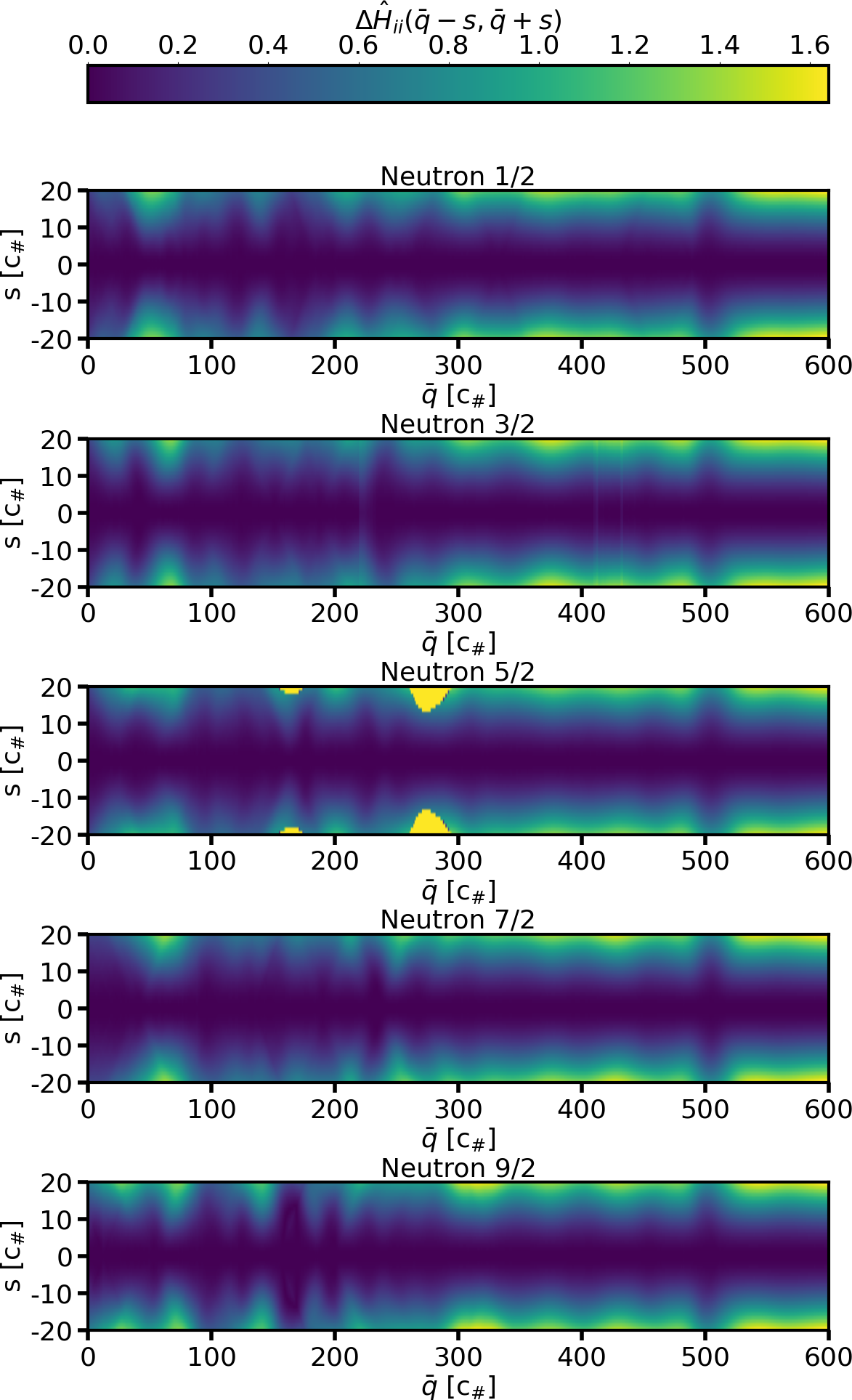}
\caption{$\Delta \hat H_{ii}$ for all neutron variational excitations as a function of $\bar q$ and $s$.}
\label{ctwo_88}
\end{figure}

The evaluation of the overlap kernels is detailed in Appendix D of the first article of the trilogy, while the matrix elements $\bra{\Phi_{i}(\bar q-s)}\hat H \ket{\Phi_i(\bar q+s)}$ are discussed in Appendix F.

FIG. \ref{ctwo_88} shows $\Delta \hat H_{ii}$ for all neutron variational excitations as a function of the center-of-mass coordinate $\bar q$ and the relative coordinate $s$, expressed in units of $c_\#$. The overall behavior is found to be consistent with that of the adiabatic reference case. One exception is observed for the neutron excitation with $\Omega=5/2$, for which, around $\bar q \simeq 270$ and for $|s| > 13$, deviations as large as 77\% are obtained. Given that these deviations occur at large values of $|s|$, they do not question the overall regularity of the corresponding Hamiltonian kernels. However, as will be discussed in the third article of the trilogy, this feature has a non-negligible impact on the dynamics, in particular on the inertia tensor. A similar behavior is also found around $c_\# \simeq 170$ for the same excitation. This issue may have a numerical origin and will be further investigated in future work.

\begin{figure}
\centering
\includegraphics[width=1.0\linewidth]{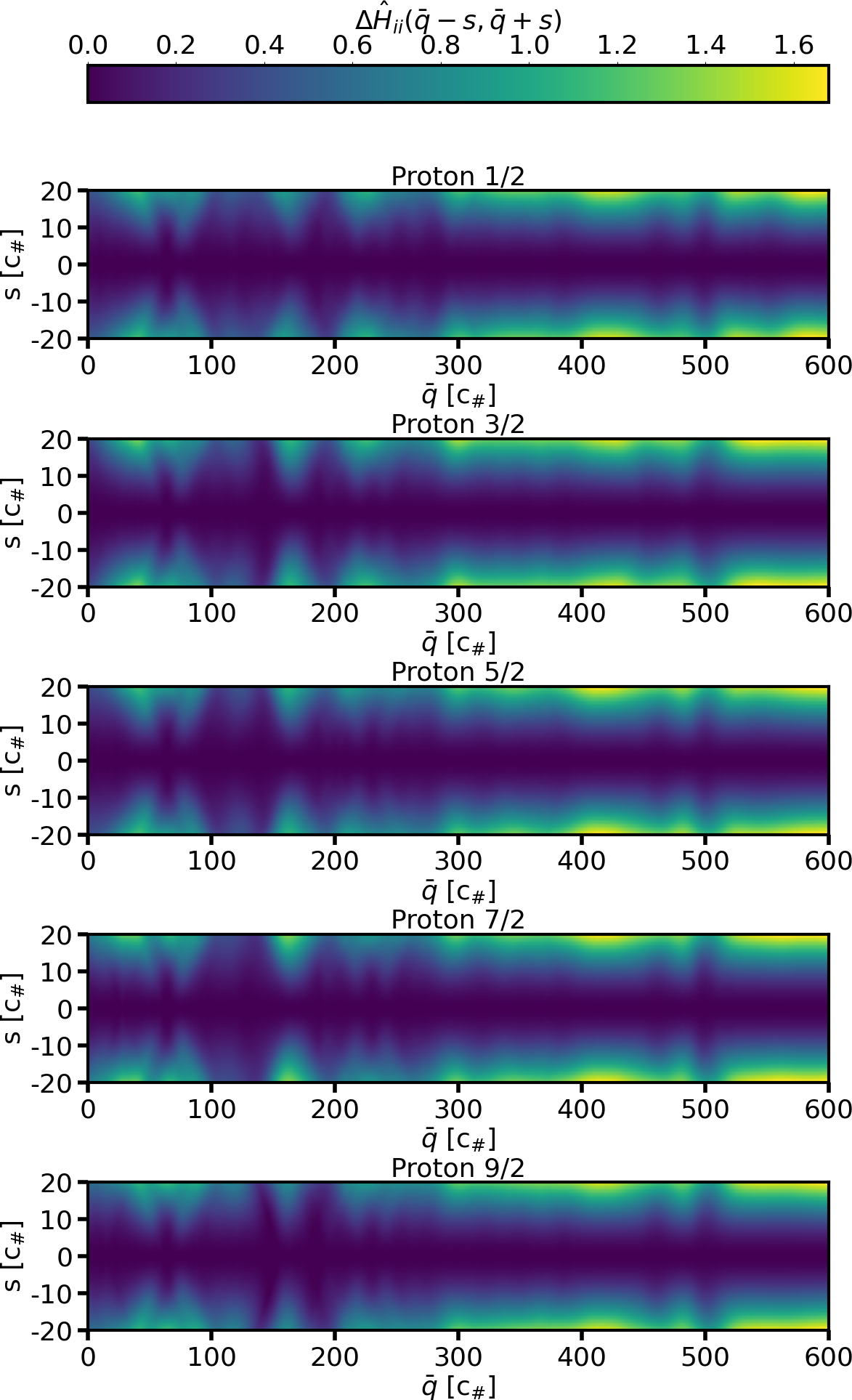}
\caption{$\Delta \hat H_{ii}$ for all proton variational excitations as a function of $\bar q$ and $s$.}
\label{ctwo_89}
\end{figure}

FIG. \ref{ctwo_89} presents the corresponding results for the proton variational excitations. For protons, the overall behavior of $\Delta \hat H_{ii}$ is very satisfactory, showing a level of regularity fully consistent with the neutron case, but without any noticeable exception.

The situation for off-diagonal Hamiltonian kernels is similar to that of the off-diagonal overlap kernels. No established benchmark exists to assess their regularity. Nevertheless, their magnitude is significantly smaller than that of the diagonal Hamiltonian kernels, suggesting that their contribution to the overall dynamics is subleading. In practice, no pathological behavior associated with these terms has been observed in the dynamical evolution. However, the numerical evaluation of the off-diagonal Hamiltonian kernels can become delicate in the limit $s \rightarrow 0$, which is addressed in the following section.

\subsubsection{Numerical instabilities in off-diagonal Hamiltonian kernels}
\label{divsol}

In practice, no particular difficulty is encountered in the evaluation of overlap and Hamiltonian kernels, except in two specific situations: Hamiltonian kernels between adiabatic states and variational excited states, and kernels between two distinct variational excited states. In both cases, the expressions involve ratios in which numerator and denominator vanish simultaneously, with the denominator approaching zero more rapidly than the numerator (quadratic behavior, see Ref.~\cite{DobaDiv}).

We initially assumed that the improvement in numerical accuracy associated with the orthogonality constraints on the overlap kernels would mitigate this issue. However, residual instabilities remain, with varying severity depending on the excitation.

\begin{figure}
\centering
\includegraphics[width=1.0\linewidth]{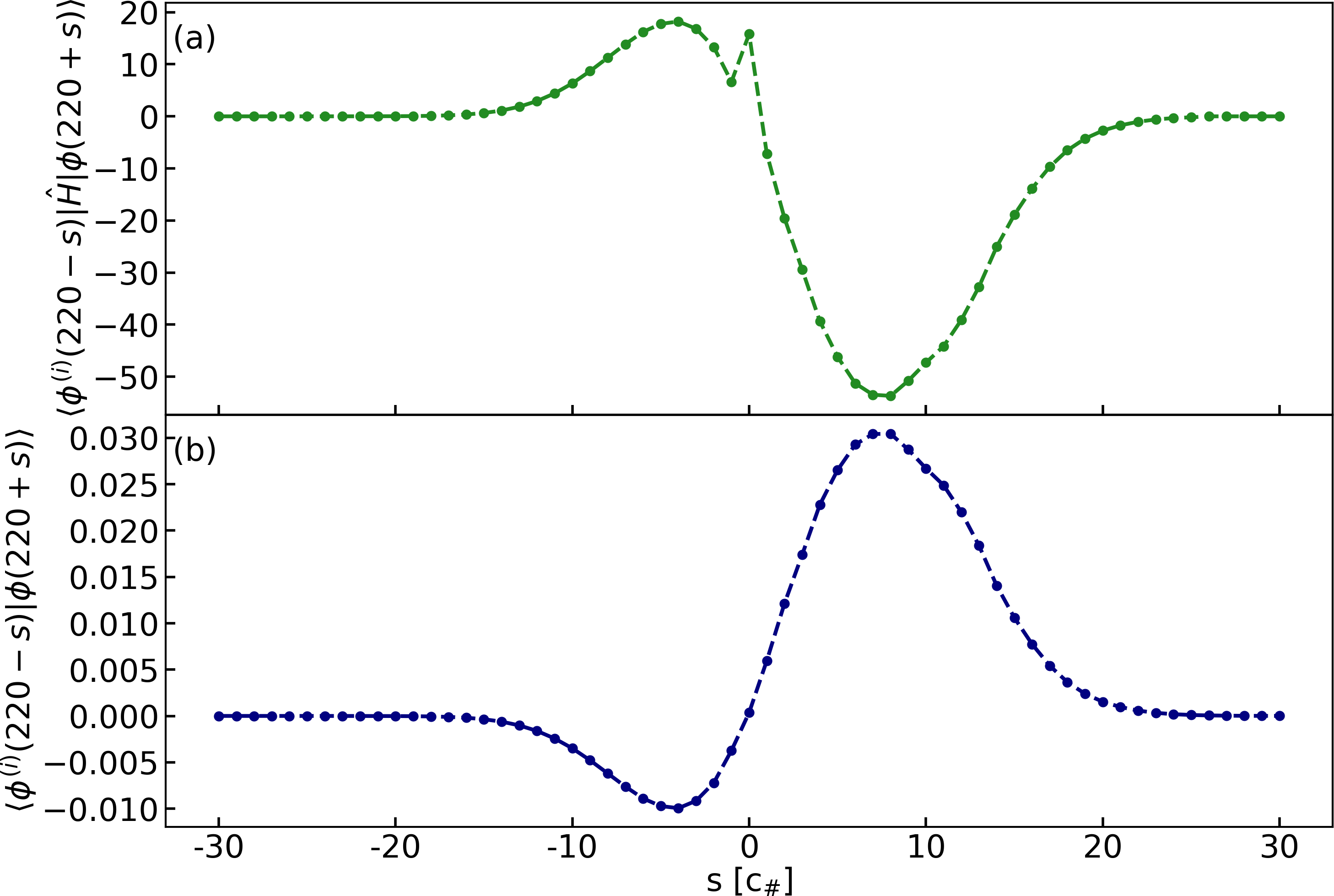}
\caption{Illustration of small Hamiltonian kernel divergences. Panel (a): Hamiltonian kernel $\bra{\Phi^{(i)}(220 - s)}\hat H \ket{\Phi(220+s)}$, with $(i)$ standing for the neutron $\Omega=1/2$ variational excitation, with respect to $s$ in c$_\#$ unit. Energies are expressed in MeV. Panel (b): Associated overlap kernel $\bra{\Phi^{(i)}(220 - s)} \ket{\Phi(220+s)}$.}
\label{ctro_136}
\end{figure}

\begin{figure}
\centering
\includegraphics[width=1.0\linewidth]{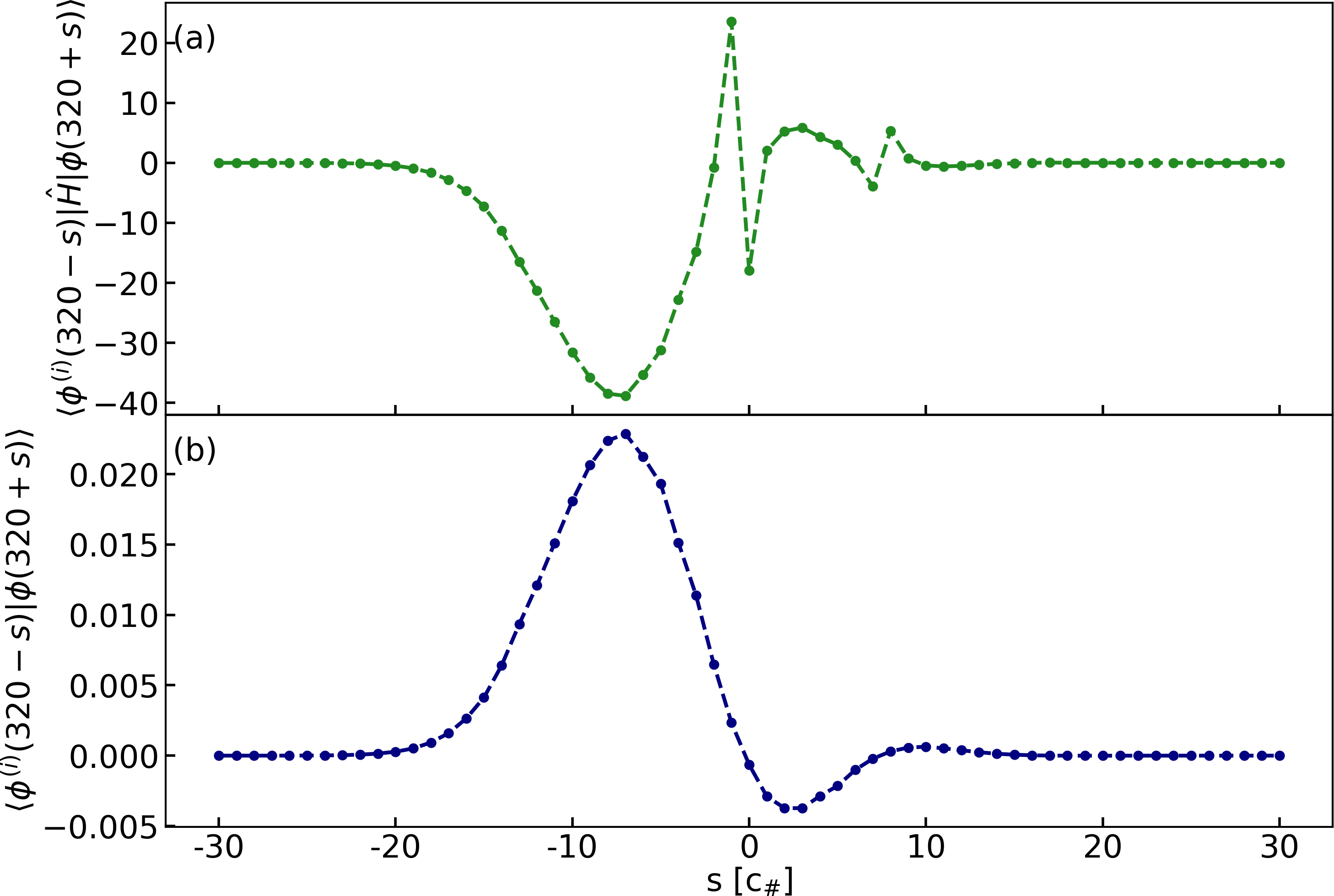}
\caption{Illustration of severe Hamiltonian kernel divergences. Panel (a): Hamiltonian kernel $\bra{\Phi^{(i)}(320 - s)}\hat H \ket{\Phi(320+s)}$, with $(i)$ standing for the neutron $\Omega=1/2$ variational excitation, with respect to $s$ in c$_\#$ unit. Energies are expressed in MeV. Panel (b): Associated overlap kernel $\bra{\Phi^{(i)}(320 - s)} \ket{\Phi(320+s)}$.}
\label{ctro_137}
\end{figure}

Figures~\ref{ctro_136} and \ref{ctro_137} illustrate two representative cases for the neutron $\Omega=1/2$ excitation. In the first case (FIG. \ref{ctro_136}), the anomaly is well localized around $s=0$ and remains moderate compared to the overall scale of the kernel. In the second case (FIG. \ref{ctro_137}), more severe deviations are observed, extending over a finite range of $s$. This behavior correlates with very small values of the associated overlap kernel in this region. In both cases, the Hamiltonian and overlap kernels remain consistent when excluding the pathological domain.


To regularize these instabilities, three prescriptions have been tested. The first, referred to as the overlap prescription, exploits the similarity between Hamiltonian and overlap kernels by replacing the Hamiltonian kernel with a weighted overlap form:
\begin{eqnarray}
\displaystyle  \bra{\Phi^{(i)}(\bar q - s)}\hat H \ket{\Phi^{{\textcolor{white}{(}}}(\bar q +s)}_{cor} = \frac{\displaystyle E^{(i)}(\bar q) + E(\bar q)}{\displaystyle 2} \nonumber \\ \displaystyle  \times \bra{\Phi^{(i)}(\bar q - s)}\ket{\Phi(\bar q +s)}
\end{eqnarray}

\begin{figure}
\centering
\includegraphics[width=1.0\linewidth]{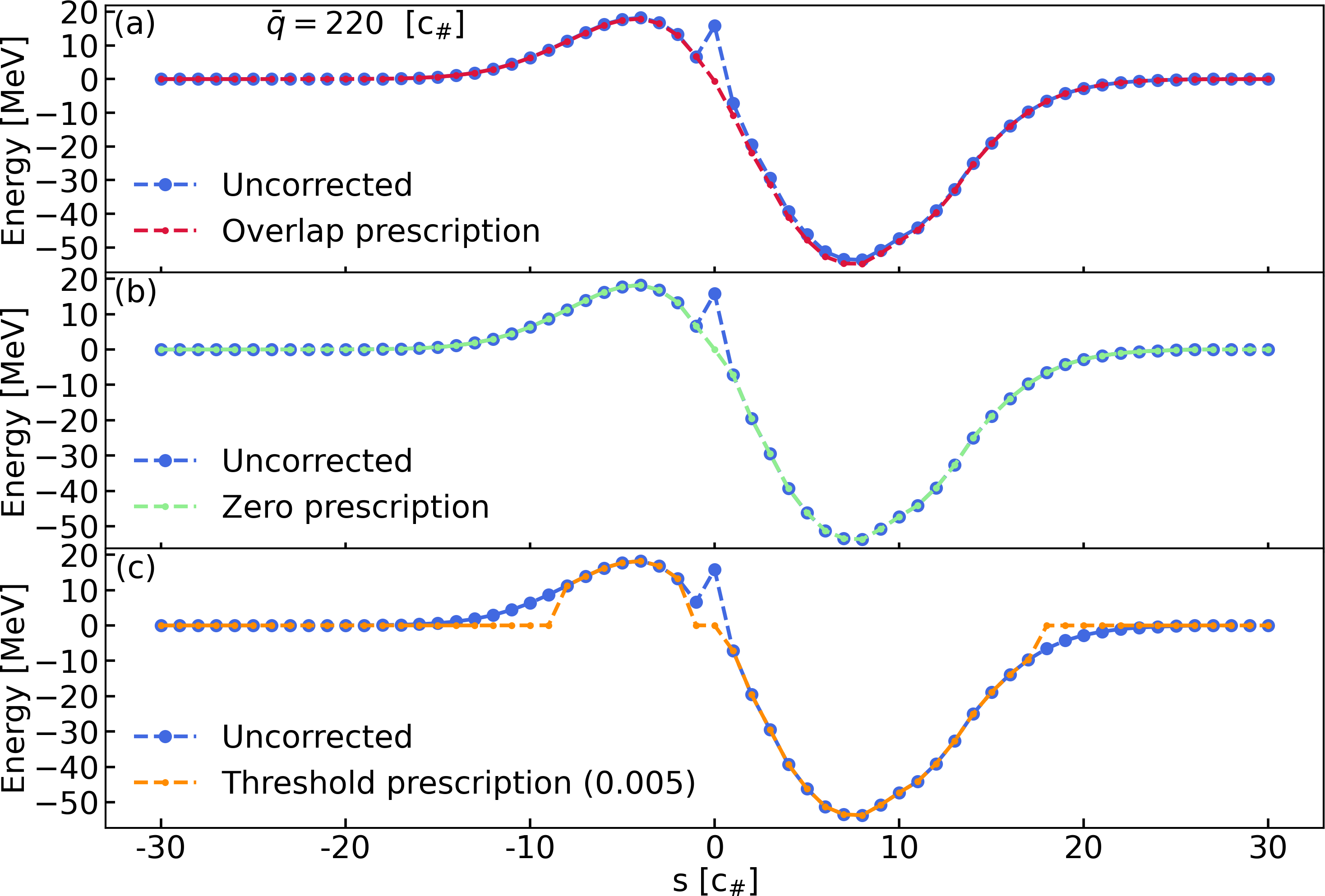}
\caption{Hamiltonian kernel prescriptions at $\bar q = 220$ [c$_\#$]. Panel (a): Overlap prescription. Panel (b): Zero prescription. Panel (c): Threshold prescription.}
\label{ctro_138}
\end{figure}

\begin{figure}
\centering
\includegraphics[width=1.0\linewidth]{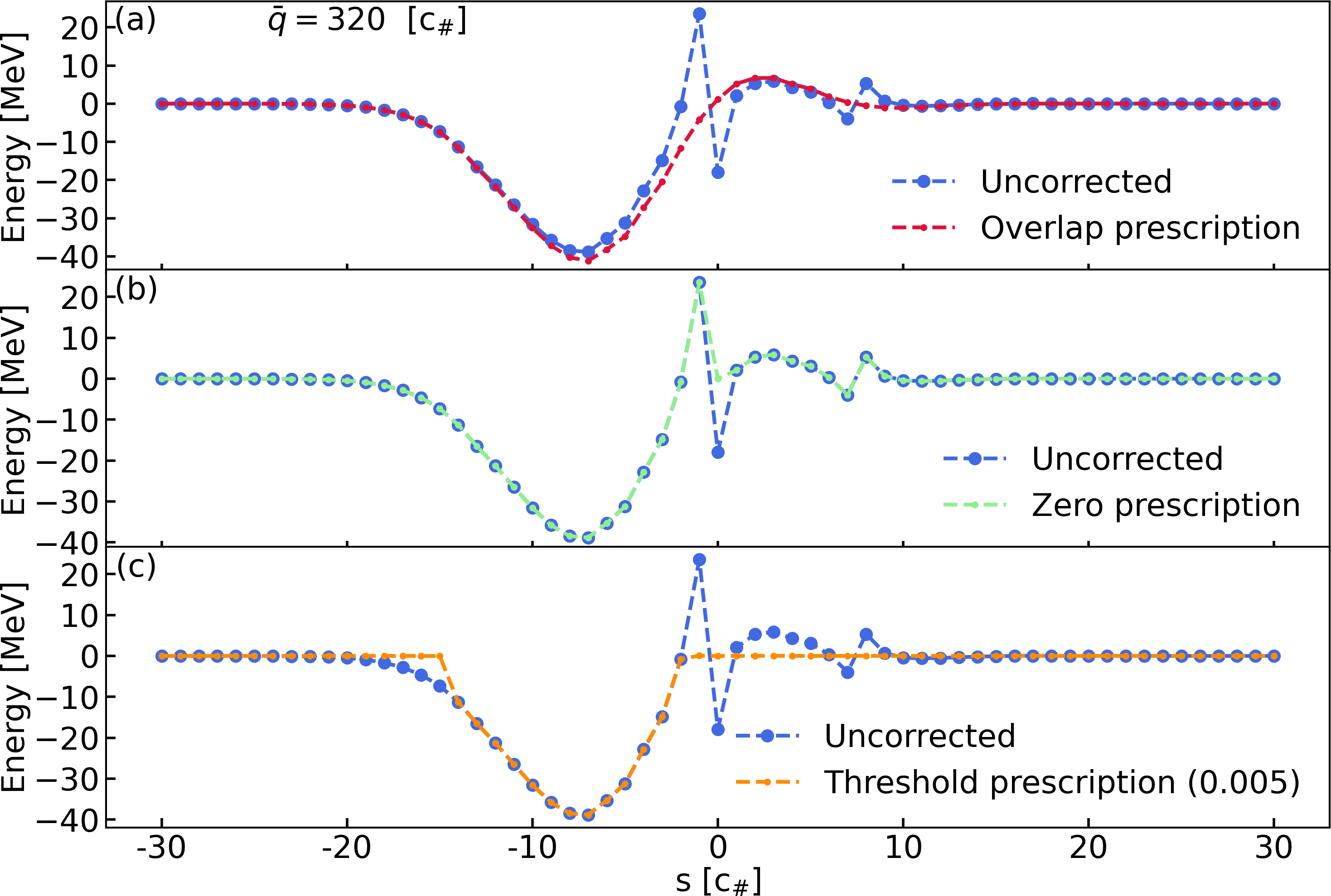}
\caption{Same as FIG.\ref{ctro_138} but for $\bar q = 320$ [c$_\#$].}
\label{ctro_139}
\end{figure}

The other two prescriptions follow Ref. \cite{DobaDiv} and rely on the idea of compensating the vanishing denominator by multiplying by the overlap kernel. The first variant, the zero prescription, corrects only the $s=0$ point:
\begin{eqnarray}
\displaystyle  \bra{\Phi^{(i)}(\bar q - s)} \hat H \ket{\Phi^{{\textcolor{white}{(}}}(\bar q +s)}_{cor} = \qquad \qquad \nonumber \\\bra{\Phi^{(i)}(\bar q - s)} \hat H \ket{\Phi^{{\textcolor{white}{(}}}(\bar q +s)} \qquad \qquad \nonumber \\  \times \left[ \delta_{s\ne 0} + \delta_{s=0} \bra{\Phi^{(i)}(\bar q - s)}\ket{\Phi(\bar q +s)} \right].
\end{eqnarray}
Given the spatial extent of the anomalies observed in FIG. \ref{ctro_137}, we also introduce a threshold prescription:
\begin{eqnarray}
 \bra{\Phi^{(i)}(\bar q - s)} \hat H \ket{\Phi^{{\textcolor{white}{(}}}(\bar q +s)}_{cor} = \qquad \qquad \nonumber \\ \bra{\Phi^{(i)}(\bar q - s)} \hat H  \ket{\Phi^{{\textcolor{white}{(}}}(\bar q +s)} \nonumber \qquad \qquad \\ \times \left[ \delta_{s>\epsilon} + \delta_{s\leq \epsilon} \bra{\Phi^{(i)}(\bar q - s)}\ket{\Phi(\bar q +s)} \right].
\end{eqnarray}  
FIGs. \ref{ctro_138} and \ref{ctro_139} compare the three prescriptions for the two representative cases, with $\epsilon = 5\times10^{-3}$. For mild instabilities, both the overlap and zero prescriptions effectively smooth the divergence, whereas the threshold prescription introduces discontinuities at the cutoff. In the more severe case, the overlap prescription yields the most stable and physically smooth behavior, while the zero prescription is insufficient and the threshold prescription partially regularizes the divergence but distorts the kernel behavior.

Based on these observations, the overlap prescription is adopted for the treatment of problematic Hamiltonian kernels. A key advantage of this approach is that it requires only diagonal Hamiltonian matrix elements within the SCIM framework, which significantly reduces the computational cost when multiple variational excitations are considered. A more systematic assessment of its accuracy is deferred to future work.

\section{Fragment properties in the excited states around the scission area of $^{240}$Pu}\label{scissionarea}

In this section, we study various properties of the different paths in the scission area which are systematically compared to the adiabatic ones. In particular, we detail le behavior of the chemical potentials, the neutron necking as well as the proton and neutron distributions of fragments.

\subsection{Chemical potentials}

\begin{figure}
\centering
\includegraphics[width=1.0\linewidth]{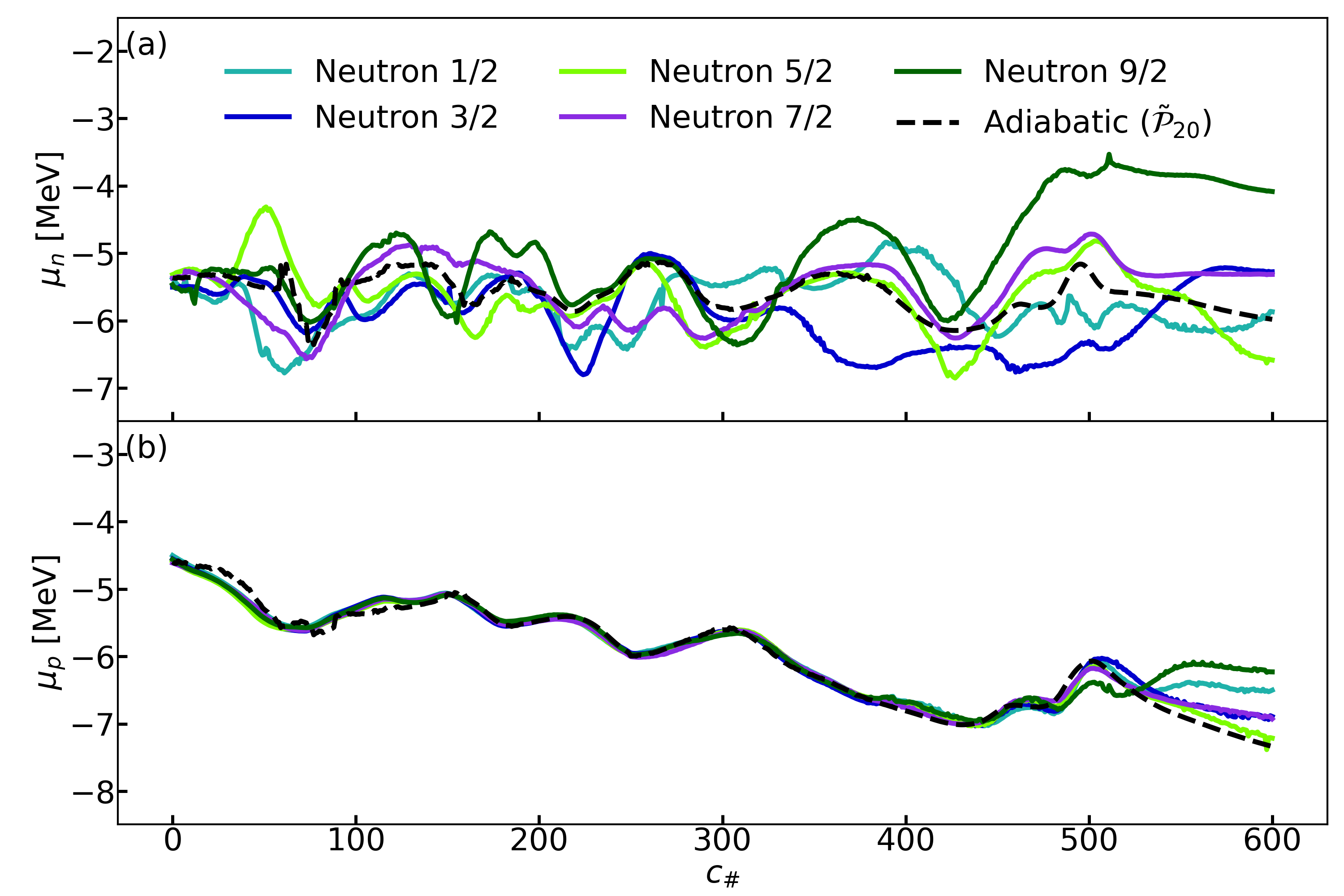}
\caption{Chemical potentials associated with five neutron variational excitations built with the $\mathcal{\tilde P^*}_{20}$ procedure on top of an adiabatic set built with the $\mathcal{\tilde P}_{20}$ procedure in the $^{240}$Pu with respect to the collective variable $c_{\#}$. Panel (a): neutron chemical potentials. Panel (b): proton chemical potentials.}
\label{ctwo_104}
\end{figure}

Building on the adiabatic analysis, we now investigate the impact of variational excitations on the neutron and proton chemical potentials. FIG. \ref{ctwo_104} shows the chemical potentials associated with the five neutron variational excitations together with the adiabatic reference states. Panel (a) displays the neutron chemical potentials, while panel (b) shows the proton ones.

The neutron variational excitations produce markedly different behaviors in the neutron chemical potentials. The characteristic adiabatic peak is preserved for the $\Omega=5/2$ and $\Omega=7/2$ excitations, although in a broader form, whereas it is barely visible for the remaining excitations. This behavior suggests that intrinsic contributions associated with the variational excitations may dominate the collective effects responsible for the adiabatic peak. 

In contrast, the proton chemical potentials remain very close to the adiabatic results up to the scission region. The subsequent spreading of the curves reflects the modifications of the particle-number distributions in the fission fragments.

\begin{figure}
\centering
\includegraphics[width=1.0\linewidth]{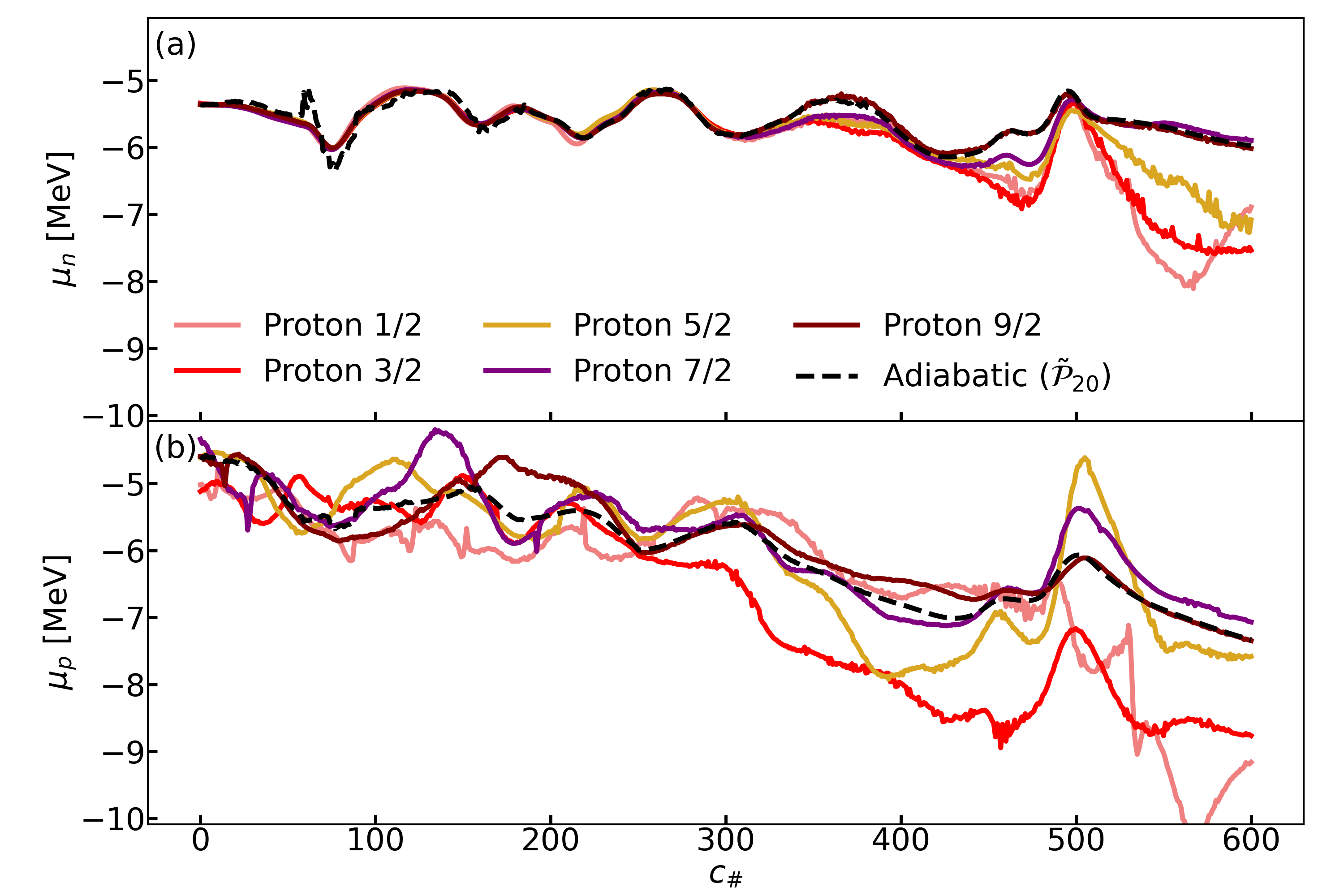}
\caption{Same as FIG. \ref{ctwo_104} but for the five proton variational excitations.}
\label{ctwo_110}
\end{figure}

FIG. \ref{ctwo_110} presents the corresponding results for the proton variational excitations. In panel (a), the neutron chemical potentials start to deviate from the adiabatic behavior earlier than in FIG. \ref{ctwo_104} (b), indicating that intrinsic modifications of comparable magnitude have a stronger impact on the proton subsystem, which contains fewer particles.

The proton chemical potentials shown in FIG. \ref{ctwo_110} (b) display a behavior analogous to that observed for the neutron chemical potentials in FIG. \ref{ctwo_104} (a). Each excitation exhibits a distinct evolution, with pronounced peaks for the $\Omega=3/2$, $5/2$, $7/2$, and $9/2$ states, while the $\Omega=1/2$ excitation shows no clear peak. These results further support the conclusion that intrinsic effects may dominate the collective effect responsible for the adiabatic peak.


\subsection{Neutron necking}

\begin{figure}
\centering
\includegraphics[width=1.0\linewidth]{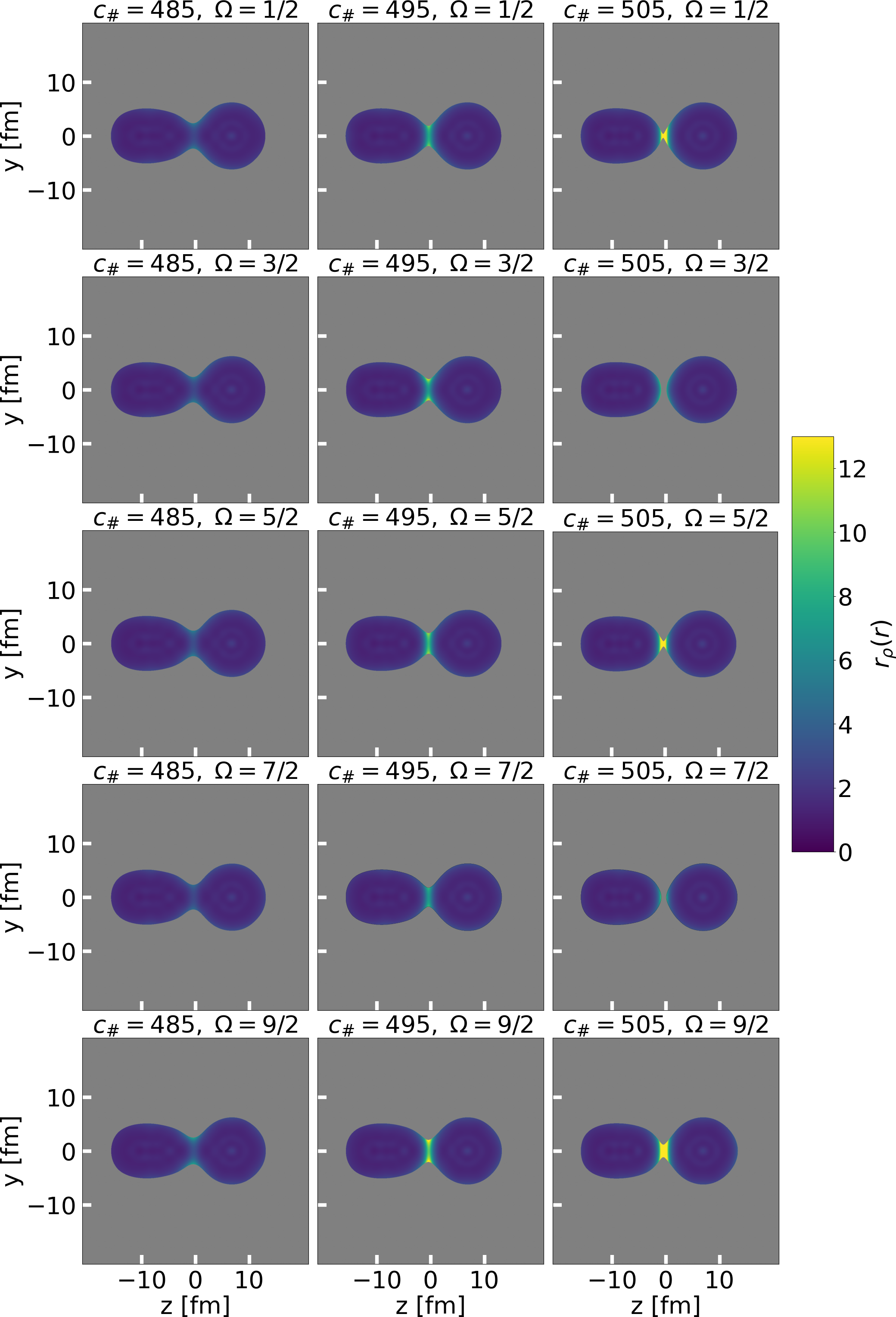}
\caption{Local neutron/proton ratio $r_\rho$ for different neutron variational excited states whose $\Omega$ ranges from 1/2 to 9/2 and labeled by $c_\# = 485$, $c_\# = 495$ and $c_\# = 505$.}
\label{ctwo_157}
\end{figure}

In this section, we investigate whether the neutron necking observed at the adiabatic level persists in the variational excitations. FIG. \ref{ctwo_157} displays the local neutron-to-proton density ratio, $r_\rho$, for the five neutron variational excitations at three representative deformations, corresponding to the three black crosses shown in FIG. 19 of the first article of the trilogy \cite{trilogy1}.

The variational excitations exhibit a markedly different behavior from the adiabatic reference. Not only are the local density ratios between the two pre-fragments significantly larger, but they also display a pronounced dependence on the excited state. For clarity, the values of $r_\rho$ shown in FIG. \ref{ctwo_157} are capped at 13. The corresponding maxima, reported in Table~\ref{ctwo_159}, exceed 20 for several excitations.

\begin{table}[htb!]
\begin{tabular}{|c|c|c|c|c|c|}
\hline &  n$_{1/2}$ & n$_{3/2}$ & n$_{5/2}$ & n$_{7/2}$ & n$_{9/2}$ \\
\hline c$_\#$ = 485 & 5.56 & 8.62 &  5.65 & 4.66 & 7.65 \\
\hline c$_\#$ = 495 & 10.40 & 12.38 & 11.30 & 8.73 & 14.22  \\
\hline c$_\#$ = 505 & 20.21 & 9.23 & 20.48 & 9.41  & 29.13 \\
\hline
\end{tabular}
\caption{Maximum values of the local ratio $r_\rho$ for the states considered in FIG.  \ref{ctwo_157}.}
\label{ctwo_159}
\end{table}

A closer inspection of the configurations at $c_\#=505$ reveals that some excited states have already separated into two fragments, whereas others remain connected according to the criterion $\rho(\mathbf{r})>5\times10^{-3}$. This behavior correlates strongly with the maximum value of $r_\rho$: the larger the peak value of the local density ratio, the later the fragment separation occurs. From a broader perspective, the systematically later fragment separation observed for the neutron variational excitations can substantially delay the scission process, an effect expected to influence the evaluation of the total kinetic energy (TKE). This observation further emphasizes the importance of incorporating intrinsic excitations into the dynamical description.

These results also provide additional insight into the coupling between the pre-fragments induced by the variational excitations. The scission point appears to constitute a sensitive indicator of this coupling strength. In particular, the $\Omega=7/2$ excitation, which undergoes the earliest separation, is also the one whose neutron chemical potentials remain closest to the adiabatic reference (FIG. \ref{ctwo_104}) and whose QP composition, characterized by $\sigma^{(2)}$ and $\sigma^{(4)}$, exhibits the greatest stability. By contrast, the $\Omega=1/2$ and $\Omega=9/2$ excitations remain connected over a larger deformation range and are associated with more atypical chemical-potential evolutions as well as larger variations in $\sigma^{(2)}$ and $\sigma^{(4)}$.

\begin{figure}
\centering
\includegraphics[width=1.0\linewidth]{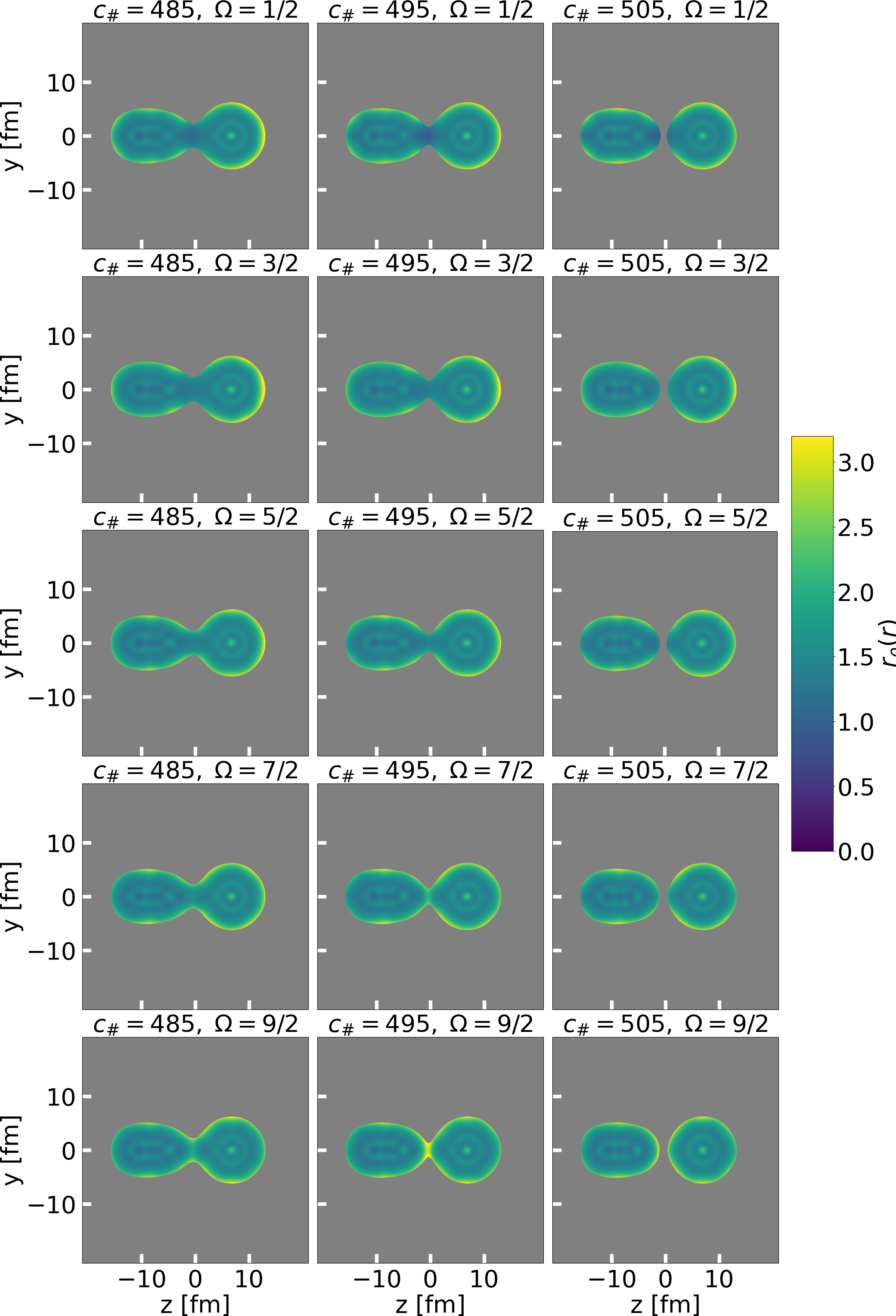}
\caption{Local neutron/proton ratio $r_\rho$ for different proton variational excited states whose $\Omega$ ranges from 1/2 to 9/2 and labeled by $c_\# = 485$, $c_\# = 495$ and $c_\# = 505$.}
\label{ctwo_158}
\end{figure}

In the context of neutron emission at scission, it is instructive to recall that, in an idealized picture of complete separation, the local density ratio $r_\rho$ would diverge. The very large values observed in the neutron variational excitations may therefore be viewed as a precursor of this behavior, although this interpretation remains qualitative at this stage.

To quantify the trends inferred from the local density ratio $r_\rho$, we now examine the evolution of the neck observable $Q_{neck}$ along the fission path \cite{QNeckWardaFirst}. One recalls that it is defined by:
\begin{eqnarray}
\hat{Q}_{neck} = exp \left( \frac{-z^2}{a^2_{neck}} \right)
\label{qneck}
\end{eqnarray}
with $a_{neck} = 1$fm.

\begin{figure}
\centering
\includegraphics[width=1.0\linewidth]{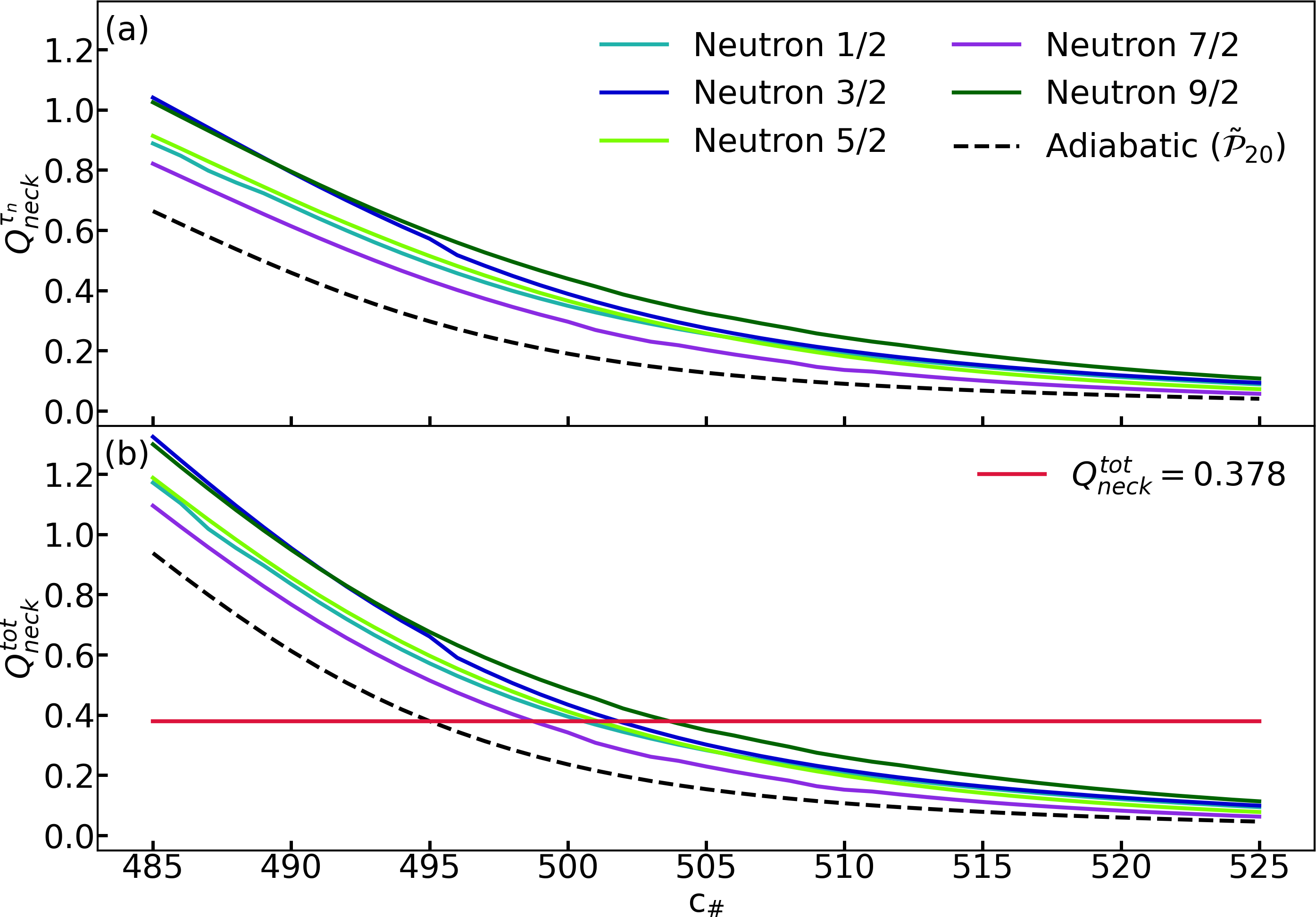}
\caption{Panel (a): Evolution of the neutron $Q_{neck}^{\tau_n}$ with respect to $c_\#$ for the five neutron variational excitations. Panel (b): Same as panel (a) but for $Q_{neck}^{tot}$.}
\label{ctwo_160}
\end{figure}

FIG. \ref{ctwo_160} (a) displays the neutron contribution $Q_{neck}^{\tau_n}$ for the five neutron variational excitations together with the adiabatic reference, while panel~(b) shows the corresponding total quantity $Q_{neck}^{tot}$. The horizontal red line indicates the value of $Q_{neck}^{tot}$ at the neutron chemical-potential peak of the adiabatic path ($c_\#=495$).

\begin{figure}
\centering
\includegraphics[width=1.0\linewidth]{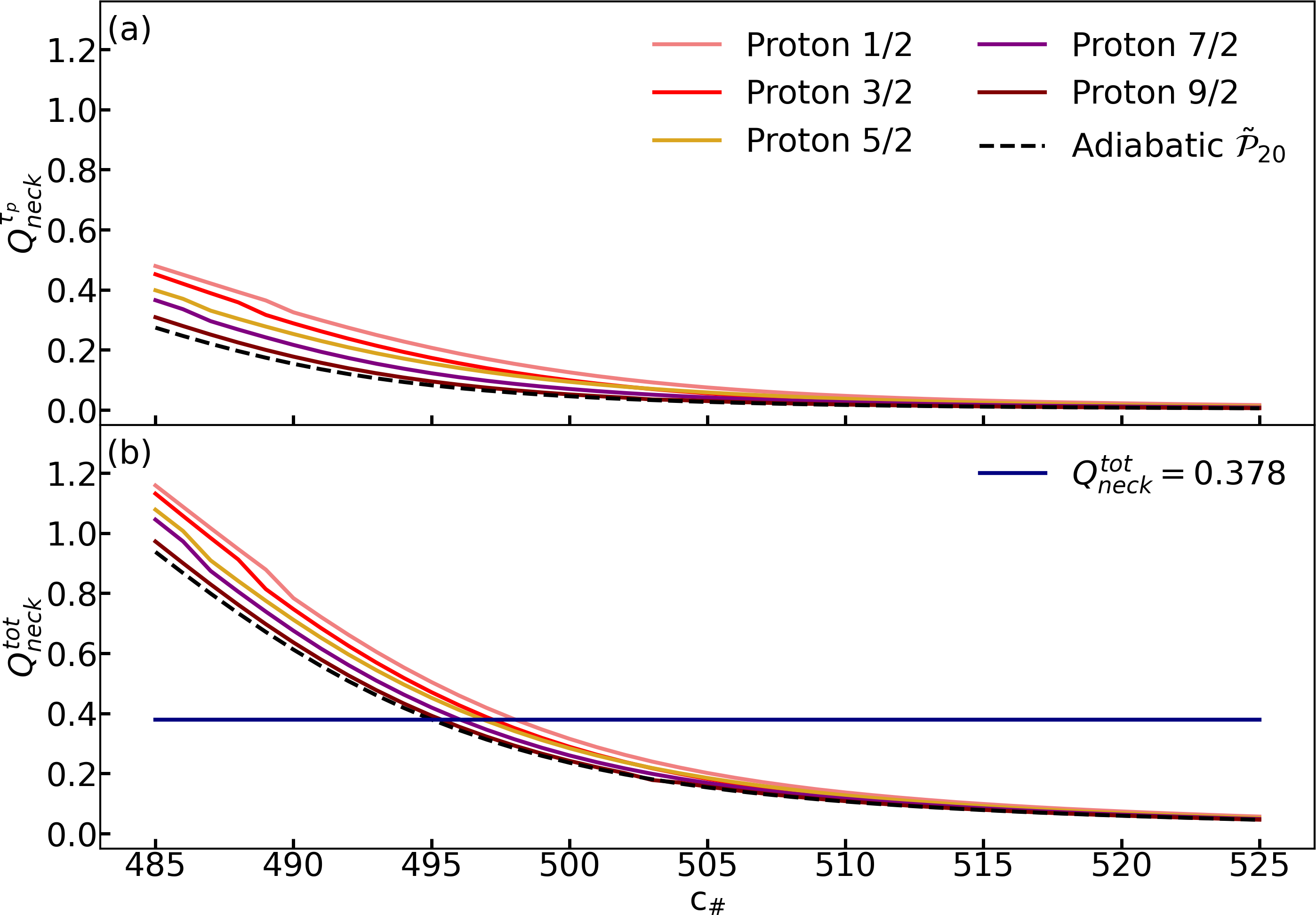}
\caption{Panel (a): Evolution of the proton $Q_{neck}^{\tau_p}$ with respect to $c_\#$ for the five proton variational excitations. Panel (b): Same but for $Q_{neck}^{tot}$.}
\label{ctwo_161}
\end{figure}

Overall, the evolution of $Q_{neck}$ confirms the trends already identified from $r_\rho$. In particular, the $\Omega=7/2$ excitation remains closest to the adiabatic reference, whereas the $\Omega=9/2$ excitation exhibits the largest deviation, consistent with its delayed scission behavior. One notable exception concerns the $\Omega=3/2$ excitation, whose evolution differs from that suggested by $r_\rho$. Since $\hat Q_{neck}$ defined in Eq. \eqref{qneck} involves a Gaussian spatial weighting, this discrepancy may reflect a specific neutron localization pattern in the neck region for this excitation. The kink observed in this curve further suggests that a more detailed analysis would be required.

We now turn to the proton variational excitations and perform a similar analysis of the local density ratio $r_\rho$. FIG. \ref{ctwo_158} shows the corresponding profiles at the same three representative deformations as in the neutron case.

Overall, the proton results exhibit a weaker sensitivity to the variational excitations than in the neutron case. In particular, no saturation effects are observed in $r_\rho$, and the spatial structure between the pre-fragments remains closer to the adiabatic configuration. A systematic reduction of the neutron-to-proton ratio in the neck region is observed compared to the adiabatic case, with a residual neutron neck still visible in the $\Omega=9/2$ excitation, and to a lesser extent in the $\Omega=7/2$ case, while it is essentially absent in the other excitations. Since the neutron content of the proton variational excitations is constrained to remain the same as in the adiabatic configuration, the observed variations can be attributed to an increased proton accumulation in the neck region.

To complement this analysis, we  compare the proton $Q_{neck}^{\tau_p}$ and total $Q_{neck}^{tot}$ in FIG. \ref{ctwo_161}. The resulting $Q_{neck}$ values for the proton variational excitations remain significantly closer to the adiabatic reference than in the neutron case. This suggests that proton variational excitations have a more limited impact on maintaining the cohesion between pre-fragments in the scission region.

\begin{figure}
\centering
\includegraphics[width=1.0\linewidth]{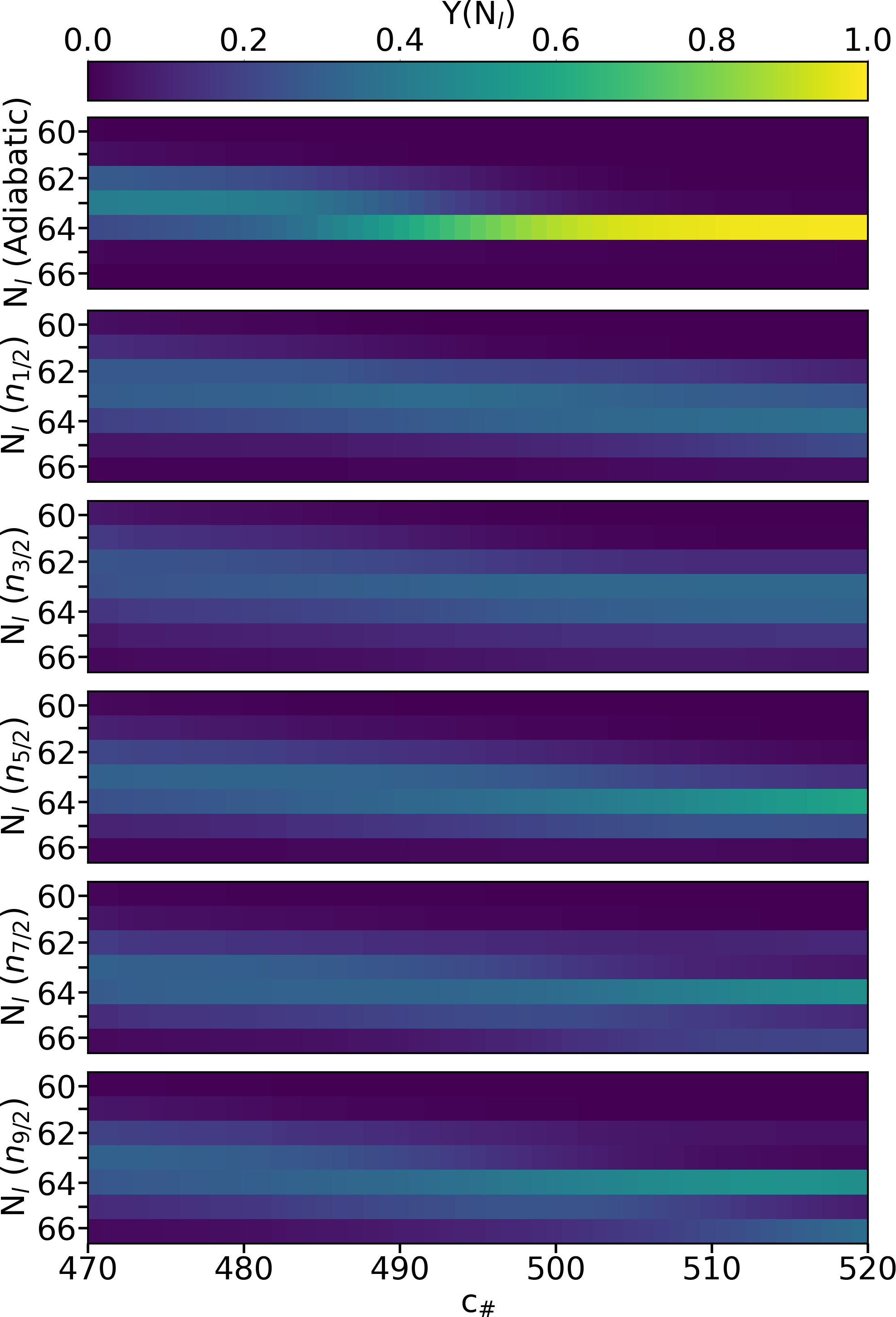}
\caption{Light fragment neutron particle number distributions of the adiabatic states and of the neutron variational excited states with respect to $c_\#$.}
\label{ctwo_177}
\end{figure}

\subsection{Proton and neutron fragment distributions}

As in the adiabatic case (see Ref.\cite{trilogy1}, section IV.B), the fragment particle number distributions for all variational excitations are obtained using the separation method of Ref. \cite{zSep}. 

\begin{figure}
\centering
\includegraphics[width=1.0\linewidth]{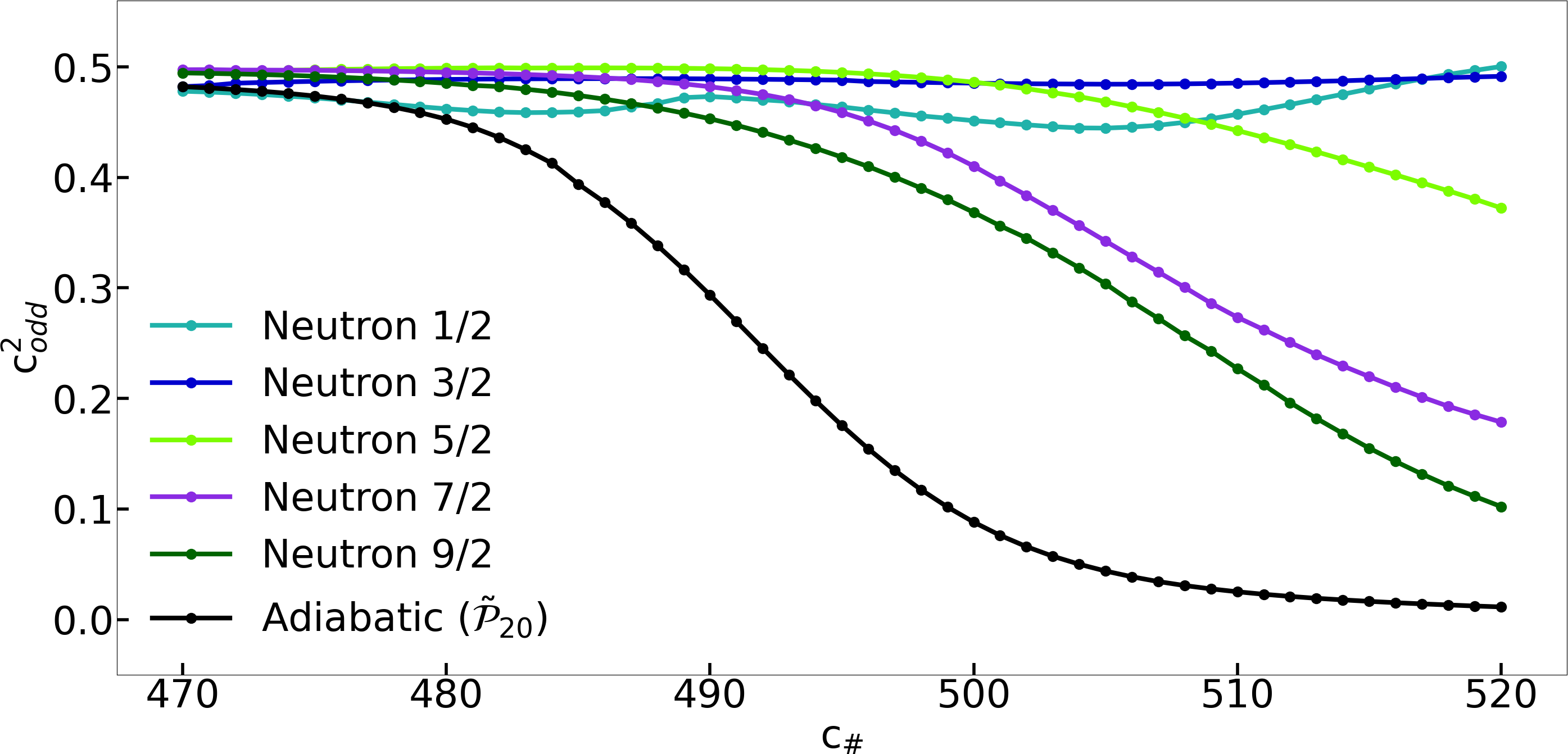}
\caption{Evolution of the odd components in the light fragment neutron particle number distributions of the adiabatic states and of the neutron variational excited states with respect to $c_\#$. }
\label{ctwo_205}
\end{figure}

We first consider the neutron distributions in FIG. \ref{ctwo_177}. While the adiabatic case exhibits a well-defined peak at $N_l=64$ beyond $c_\#=495$, the distributions associated with the neutron variational excitations are systematically broader and display non-negligible odd components. The latter are quantified in FIG. \ref{ctwo_205} through the squared norm $c_{\mathrm{odd}}^2$, which is significantly enhanced compared to the adiabatic reference. These features indicate that the variational excitations capture also the pair-breaking mechanism.

\begin{figure}
\centering
\includegraphics[width=1.0\linewidth]{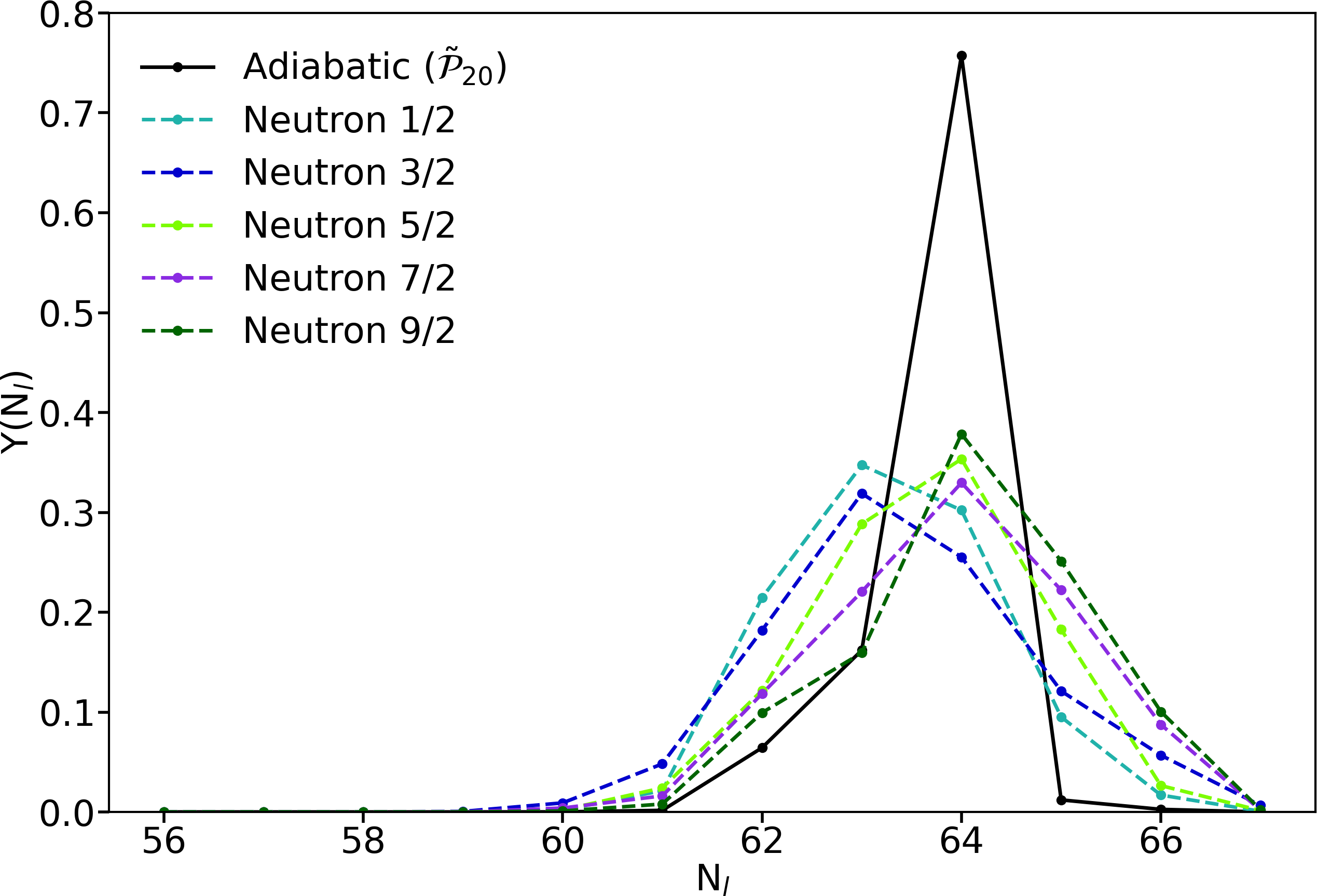}
\caption{Light fragment neutron particle number distributions of the neutron variational excited states at scission. The adiabatic one is also indicated in black.}
\label{ctwo_178}
\end{figure}

Besides, one should note that fragment distributions and mean particle numbers may still evolve with $c_\#$ even beyond scission, leading to a weak “communicating vessels” effect. For this reason, we focus on the distributions evaluated near scission. These are illustrated in FIG. \ref{ctwo_178}, where the neutron distributions are shown at $c_\#=495$. The adiabatic case remains sharply peaked at $N_l=64$, whereas the variational excitations exhibit broader profiles, in some cases dominated by odd components (e.g. $\Omega=3/2$). For comparison, experimental neutron yields show a weak peak around $N_l=60$ for the light fragment~\cite{Yield}.

Overall, neutron variational excitations significantly broaden the fragment distributions, thereby improving the description compared to the adiabatic TDGCM, which is known to produce overly narrow yields.\\

\begin{figure}
\centering
\includegraphics[width=1.0\linewidth]{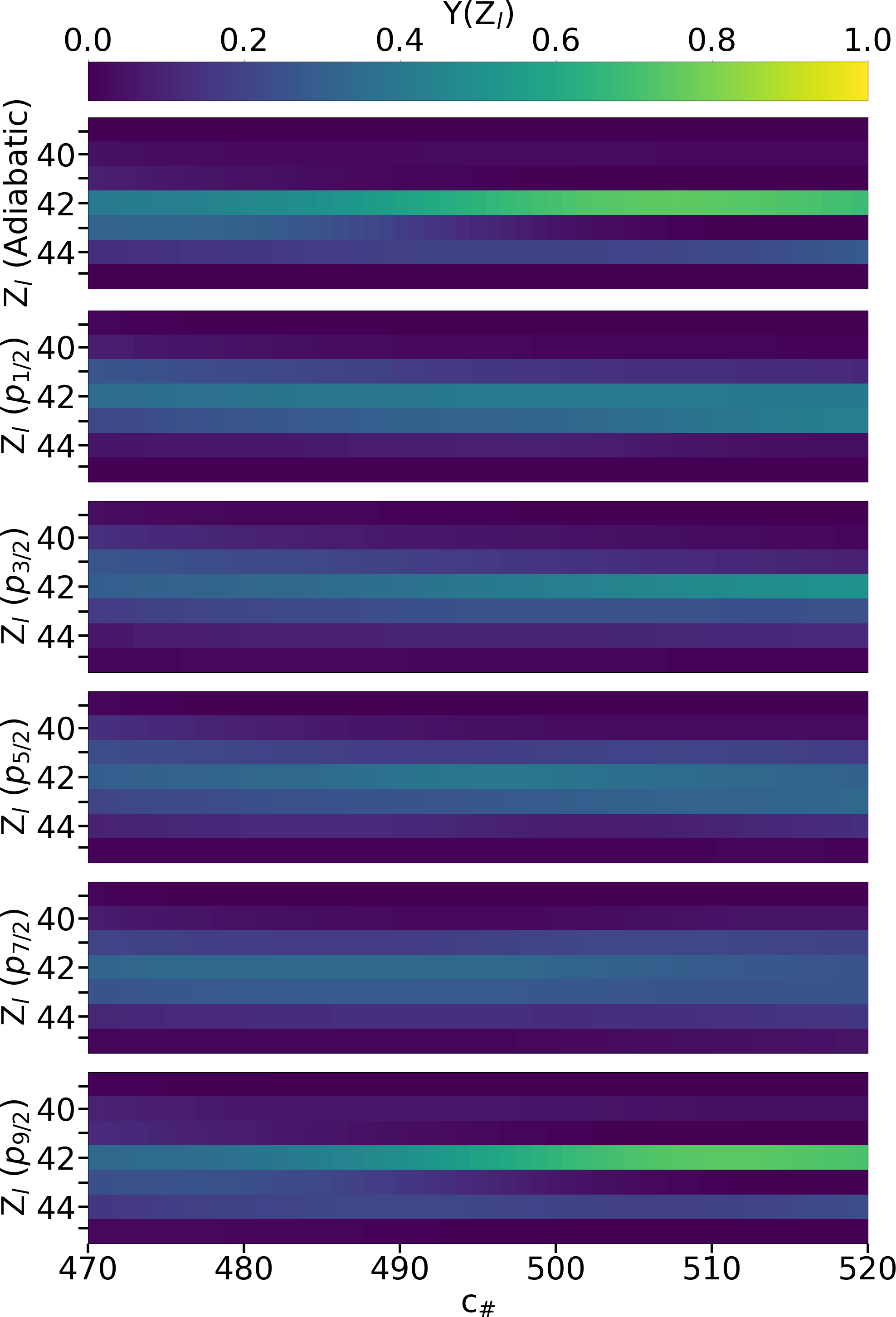}
\caption{Light fragment proton particle number distributions of the proton variational excited states with respect to $c_\#$. The adiabatic one has also been indicated.}
\label{ctwo_179}
\end{figure}

We now turn to proton distributions, illustrated in FIG. \ref{ctwo_179}. At the adiabatic level near scission, we observe two dominant contributions in the even-$Z$ sector, at $Z_l=42$ and $Z_l=44$ (in contrast, experimental charge yields exhibit a peak at $Z_l=40$ with a secondary peak at $Z_l=42$ for the light fragment), together with a clear odd-even staggering. This phenomenon reflects the strong proton pairing correlations and is absent in the neutron case.
Most proton variational excitations suppress the staggering, as expected from pair-breaking effects. A notable exception is the $\Omega=9/2$ excitation, whose distribution remains close to the adiabatic one. This is consistent with its previously identified properties, including near-adiabatic values of $\sigma^{(2)}$ and $\sigma^{(4)}$, chemical potentials, and neutron necking indicators. We therefore interpret this state as a representative example of an excitation that weakly couples the pre-fragments and is mainly localized in one of them.

\begin{figure}
\centering
\includegraphics[width=1.0\linewidth]{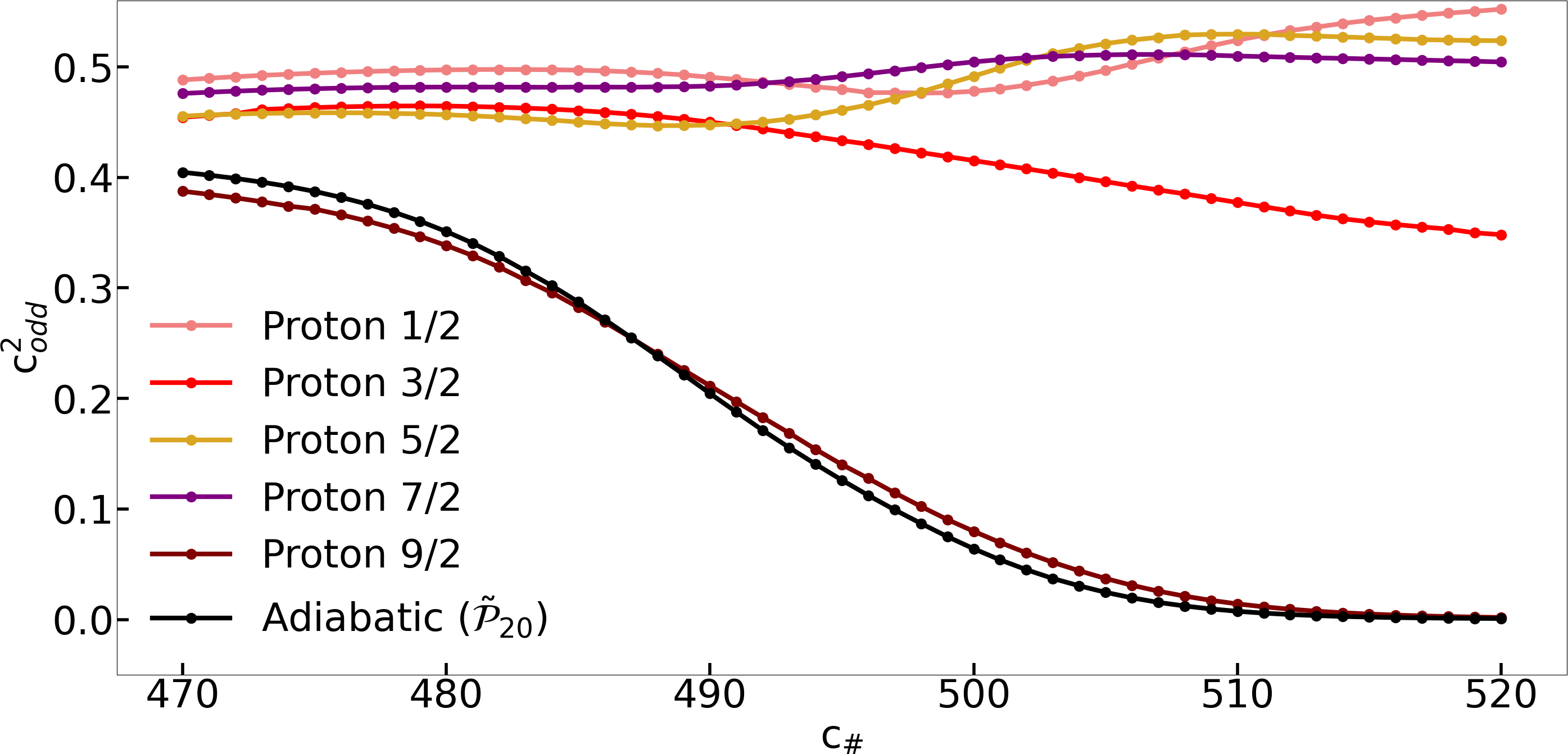}
\caption{Evolution of the odd components in the light fragment proton particle number distributions of the proton variational excited states with respect to $c_\#$. The adiabatic one has also been indicated.}
\label{ctwo_206}
\end{figure}

As in the neutron case, FIG. \ref{ctwo_206} shows the evolution of the odd components $c_{{odd}}^2$ for both adiabatic and proton variational states as a function of $c_\#$. The magnitude of $c_{{odd}}^2$ again provides a qualitative measure of pair-breaking effects and, indirectly, of the coupling between pre-fragments.

\begin{figure}
\centering
\includegraphics[width=1.0\linewidth]{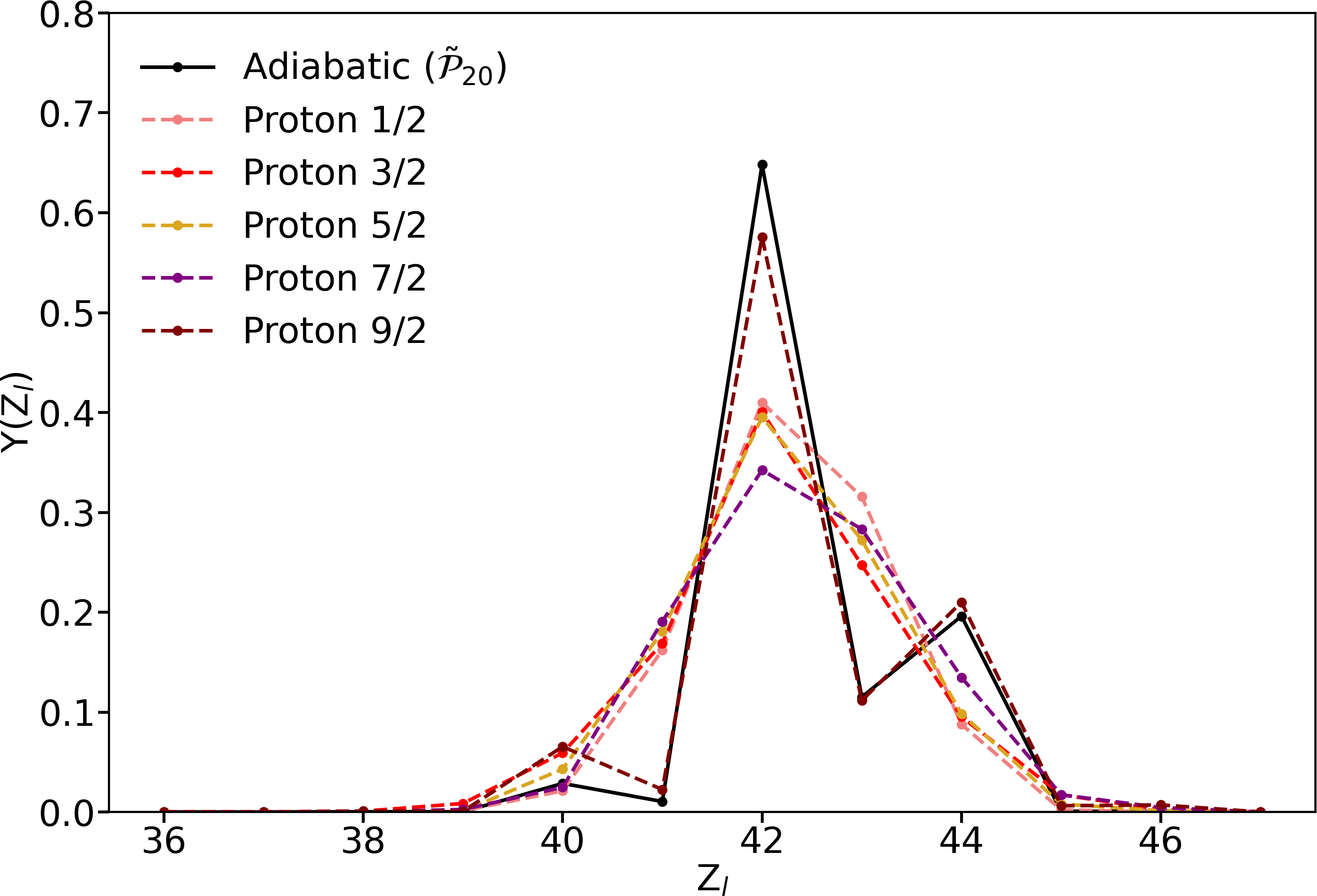}
\caption{Light fragment proton particle number distributions of the adiabatic state and of the proton variational excited states at scission.}
\label{ctwo_180}
\end{figure}

FIG. \ref{ctwo_180} displays the proton particle number distributions at scission ($c_\#=495$). The adiabatic state exhibits a pronounced odd-even staggering, whereas the variational excitations generally show broader distributions with reduced staggering, except for the $\Omega=9/2$ case, which closely follows the adiabatic behavior.

If all configurations contributing to fission preserved the strong staggering seen at the adiabatic level, the resulting sawtooth structure in the charge yields would be significantly more pronounced than in experimental data. This highlights the role of intrinsic excitations in the dynamics: they not only broaden the distributions but also reduce the excessive proton odd-even staggering, thereby improving the agreement with experimental charge yields. This aspect will be further investigated in the third article of the trilogy, devoted to dynamical effects.

\section{Conclusions and perspectives}\label{conclusion}

In this second article of the trilogy, we have addressed the severe limitations associated with pure 2QP excitations, which hinder the direct application of the SCIM method. To overcome these difficulties, we have introduced a new protocol, the \enquote{Continuous Deflation} method, which generates variational excited states. This construction relies on overlap constraints supplemented by orthogonality and continuity conditions.

We have constructed ten different excited states along the one-dimensional asymmetric fission path of $^{240}$Pu by enforcing orthogonality within a selected $\Omega$-block associated with a fixed isospin. These excitations have been smoothly propagated in deformation from the ground state up to scission and beyond. The resulting states are mixtures dominated by 2QP and 4QP configurations and exhibit clear correlations between their microscopic structure, purity indicators, and excitation energies. This framework provides access to a detailed analysis of fragment properties and observables, including chemical potentials, neutron necking, fragment mass and charge distributions, and odd components of the HFB wave function in the scission region.

As in the adiabatic case, discussed in the first paper of the trilogy \cite{trilogy1}, several perspectives remain open. A first natural extension is the construction of two-dimensional potential energy surfaces, which requires the definition of a second orthogonal collective coordinate. Another promising direction is the systematic generation of additional excited states using the \enquote{Deflation} and \enquote{Continuous Deflation} procedures, which is currently under investigation. Finally, the present work has identified a class of excited states suitable for the SCIM framework. It would be valuable to determine whether more general classes of excited states can also be consistently employed, thereby improving the applicability of the SCIM framework. \\ \\

\noindent \textbf{Acknowledgment}: N.P and P.C. would like to thank J.F. Berger
for his kindness throughout this work. N.P. dedicates this first application 
of the SCIM approach to the memory of D. Gogny. The work of W.Y. was supported by the U.S. Department of Energy, Office of Science, Office of Nuclear Physics under the contract No.
DE-AC02- 05CH11231 (LBNL). The work of L.M.R. is supported by Spanish Agencia Estatal de Investigacion 
(AEI) of the Ministry of Science and Innovation under Grant No. 
PID2024-159559NB-C21.

\bibstyle{apsrev4-2}

\bibliography{static2_SCIM_final}
\end{document}